\documentclass[%
nofootinbib,
superscriptaddress,
reprint,
amsmath,amssymb,
aps]{revtex4-2}

\usepackage{graphicx}% Include figure files
\usepackage{dcolumn}% Align table columns on decimal point
\usepackage{bm}% bold math
\usepackage{hyperref}% add hypertext capabilities
\usepackage[utf8]{inputenc}
\usepackage[brazilian]{babel}
\usepackage[T1]{fontenc}
\usepackage{amsthm}
\usepackage{amsmath}
\usepackage{color}
\usepackage{subfigure}
\usepackage{enumerate}
\usepackage{epigraph}
\usepackage{hyperref}

\usepackage{mathtools}

\usepackage{etoolbox}
\makeatletter
\patchcmd{\epigraph}{\@epitext{#1}}{\itshape\@epitext{#1}}{}{}
\makeatother

\newcommand{\ket}[1]{\left|{#1}\right>}
\newcommand{\bra}[1]{\left<{#1}\right|}
\newcommand{\braket}[2]{\left<{#1}|{#2}\right>}

\addto\captionsbrazilian{}

\usepackage[table]{xcolor}
\newcolumntype{s}{>{\columncolor[HTML]{AAACED}} p{3cm}}
\arrayrulecolor[HTML]{DB5800}

\begin{document}

\title{\textit{Quantum Finance}: um tutorial de computação quântica aplicada ao mercado financeiro}

\author{Askery Canabarro}
\email[Autor correspondente: ]{askery@gmail.com}
\affiliation{Instituto Internacional de Física, Universidade Federal do Rio Grande do Norte, 59078-970, P. O. Box 1613, Natal-RN, Brasil}
\affiliation{Grupo de F\'isica da Mat\'eria Condensada, N\'ucleo de Ci\^encias Exatas - NCEx, Campus Arapiraca, Universidade Federal de Alagoas, 57309-005, Arapiraca-AL, Brasil}
\author{Taysa M. Mendonça}
\affiliation{Instituto Internacional de Física, Universidade Federal do Rio Grande do Norte, 59078-970, P. O. Box 1613, Natal-RN, Brasil}
\author{Ranieri Nery}
\affiliation{Instituto Internacional de Física, Universidade Federal do Rio Grande do Norte, 59078-970, P. O. Box 1613, Natal-RN, Brasil}
\author{George Moreno}
\affiliation{Instituto Internacional de Física, Universidade Federal do Rio Grande do Norte, 59078-970, P. O. Box 1613, Natal-RN, Brasil}
\author{Anton S. Albino}
\affiliation{Latin American Quantum Computing Center/Senai Cimatec, Salvador-BA, Brasil}
\author{Gleydson F. de Jesus}
\affiliation{Latin American Quantum Computing Center/Senai Cimatec, Salvador-BA, Brasil}
\author{Rafael Chaves}
\email[Autor correspondente: ]{rchaves@iip.ufrn.br}
\affiliation{Instituto Internacional de Física, Universidade Federal do Rio Grande do Norte, 59078-970, P. O. Box 1613, Natal-RN, Brasil}
\affiliation{School of Science and Technology, Federal University of Rio Grande do Norte, 59078-970 Natal, Brazil}

\bigbreak 
\begin{abstract}

Antes restrita a uma área de fronteira da Física, a computação quântica é uma das áreas que mais tem crescido atualmente, justamente por suas aplicações tecnológicas em problemas de otimização, aprendizagem de máquina, segurança da informação e simulações. O objetivo deste artigo é introduzir os fundamentos da computação quântica, tendo como foco um algoritmo quântico promissor e sua aplicação a um problema do mercado financeiro. Mais especificamente, discutimos o problema da otimização de portfólio usando o \textit{Quantum Approximate
Optimization Algorithm} (QAOA). Não somente descrevemos os principais conceitos envolvidos, mas também consideramos exemplos simples práticos, envolvendo ativos financeiros disponíveis na bolsa brasileira, com códigos, tanto clássicos quanto quânticos, dísponíveis livremente em um Jupyter Notebook. Também analisamos em detalhes a qualidade das soluções de otimização combinatória de portfólio por meio do QAOA usando o simulador quântico ATOS QLM do SENAI/CIMATEC.

\begin{description}
\item[Palavras-chave] Computação Quântica, Mercado Financeiro, Finança Quântica.
\end{description}

\bigbreak 

Previously only considered a frontier area of Physics, nowadays quantum computing is one of the fastest growing research field, precisely because of its technological applications in optimization problems, machine learning, information security and simulations. The goal of this article is to introduce the fundamentals of quantum computing, focusing on a promising quantum algorithm and its application to a financial market problem. More specifically, we discuss the portfolio optimization problem using the \textit{Quantum Approximate
Optimization Algorithm} (QAOA). We not only describe the main concepts involved but also consider simple practical examples, involving financial assets available on the Brazilian stock exchange, with codes, both classic and quantum, freely available as a Jupyter Notebook. We also analyze in details the quality of the combinatorial portfolio optimization solutions through QAOA using SENAI/CIMATEC's ATOS QLM quantum simulator.

\begin{description}
\item[Keywords] Quantum Computing, Financial Market, Quantum Finance.  
\end{description}
\end{abstract}

\maketitle
\tableofcontents

\section{Introdução}
\label{sec:intro}

Desenvolvida ao longo das primeiras décadas do século 20 para explicar fenômenos na escala atômica da matéria, a teoria quântica mudou radicalmente os fundamentos da Física \cite{kumar2008quantum}. Até o advento da teoria quântica, sempre assumimos que os átomos e outros constituintes microscópicos tivessem sua posição e velocidade muito bem definidas. As probabilidades refletiam apenas nosso desconhecimento destas quantidades. Mas ao contrário da Física Clássica, a incerteza na Física Quântica é fundamental. Mesmo que saibamos tudo que há para saber sobre um elétron -- ou seja, a sua função de onda -- só podemos fazer predições probabilísticas sobre suas propriedades. Ao contrário do que estávamos acostumados até então, o papel da medição passou a ter um papel central, algo que na interpretação mais ortodoxa da mecânica quântica, conhecida como interpretação de Copenhangen, pode ser resumido nas palavras de Pascual Jordan: ``\emph{Observações não somente perturbam o que vai ser medido, elas o produzem [...] Nós compelimos o elétron a assumir uma posição bem definida [...] Nós próprios somos os responsáveis por produzir os resultados das medições que observamos}''. Ou como disse Asher Peres certa vez: ``\emph{Experimentos não realizados não tem resultados}''.

Mas nem todos no entanto concordaram com essa nova visão, certamente radical. Em particular, em 1935, Albert Einstein, Boris Podolsky e Nathan Rosen argumentaram que mecânica quântica apesar de correta -- afinal ela prediz com total acurácia as probabilidades de observação em experimentos realizados -- seria uma teoria incompleta \cite{EPR1935}. No “paradoxo de EPR”, como viria a ficar conhecido esse artigo, o chamado emaranhamento quântico, inicialmente identificado por Erwin Schrödinger \cite{Schrodinger1935}, tem um papel fundamental. Mas ao contrário do que o trio EPR poderia supor, esse emaranhamento é de fato uma característica fundamental da natureza, conforme demonstrado pelo teorema de Bell \cite{bell1964einstein} e seus variados testes experimentais \cite{hensen2015loophole,giustina2015significant} que provam de maneira categórica a inadequação do realismo local invocado por Einstein para chegar ao seu paradoxo.

Apesar de questões conceituais da mecânica quântica continuarem mais atuais do que nunca, a chamada primeira revolução quântica possibilitou avanços nas mais diversas áreas e o desenvolvimento de variadas tecnologias. A moderna teoria da computação, formulada por Alan Turing \cite{Turing1936}, por exemplo, certamente não teria tido o mesmo impacto se não fosse pelo processo de miniaturização permitido pelos transistores e outras multitudes de componentes eletrônicos que operam de acordo com as leis quânticas. Alguns anos mais tarde, em 1948, seria a vez de Claude Shannon definir matematicamente o conceito de informação \cite{Shannon1948}. Shannon não somente definiu a unidade básica de informação, o dígito binário, ou bit, como também estabeleceu a teoria que quantifica os recursos físicos necessários para armazenar, comunicar e recuperar os dados de uma fonte de informação (ou seja, o número de bits), mesmo na presença de ruído no canal. Novamente sem a quântica, neste caso responsável pelo desenvolvimento do \textit{laser} e dispositivos capazes e armazenar grandes quantidades de informação, dificilmente estaríamos a nos indagar sobre a atual era da informação.

%ele chegou a um modelo abstrato de computação conhecido como Máquina de Turing \cite{Turing1936}. Assim como os computadores modernos, a Máquina de Turing opera com um número mínimo de símbolos e instruções para realizar operações lógicas.

%De fato, o hardware de nossas máquinas, ao menos parcialmente, é quântico. O software de nossos computadores, no entanto, ainda é clássico, não muito distinto da máquina de Turing \cite{Turing1936} que deu início à ciência da computação.

%De fato, hoje em dia, estados emaranhados estão no cerne do que chamamos de a segunda revolução quântica, no qual as propriedades quânticas são utilizadas como recursos em variadas aplicações no processamento de informação, desde formas mais eficientes e seguras de se comunicar, até a metrologia quântica e computação quântica.}

%\textcolor{blue}{A teoria da computação foi formulada em 1935 por Alan Turing, ele chegou a um modelo abstrato de computação conhecido como Máquina de Turing \cite{Turing1936}. Assim como os computadores modernos, a Máquina de Turing opera com um número mínimo de símbolos e instruções para realizar operações lógicas.}

Não obstante essas revoluções primordiais, foi somente no ínicio da década de 1980 que as reais possibilidades abertas pela teoria quântica no processamento da informação começariam a tomar forma. Em 1982, Richard Feynman \cite{Feynman1982} mostrou que devido às propriedades do espaço de Hilbert no qual a informação quântica é codificada, nenhuma máquina clássica seria capaz de simular fenômenos quânticos em larga escala sem que um fator exponencial fosse introduzido em seu desempenho. Como pode ser lido em mais detalhes na edição especial da RBEF \cite{Caldeira2018} sobre o legado de Feynman, ele propôs então que apenas um “simulador quântico universal” seria capaz de fazê-lo eficientemente. Com trabalhos tais como de Benioff \cite{Benioff1980, Benioff1982}, propondo uma versão quântica da máquina de Turing,  e de Deutsch \cite{Deutsch1985}, que construiu uma situação matemática onde as características próprias dos sistemas quânticos foram usadas explicitamente para resolver um problema com funções binárias, generalizando a pedra fundamental da computação para o regime quântico, tivemos o primeiro vislumbre da computação quântica.

Em paralelo, o pioneirismo de Bennett e Brassard ao proporem a criptografia quântica \cite{BB84} e a descoberta da não-clonagem quântica por Wootters e Zurek \cite{Wootters1982} trouxeram à tona possibilidades antes impossíveis no limite clássico. Com a descoberta dos primeiros algoritmos quânticos práticos com vantagens computacionais, tais como o de Shor \cite{shor} para fatoração em número primos e o de Grover \cite{groverStandard} para a busca em um banco de dados, foi mostrou em definitivo o potencial da informação quântica. Desde então, a chamada segunda revolução quântica amadureceu rapidamente em um campo de pesquisa multidisciplinar, com efeitos de longo alcance tanto em ciência fundamental quanto em aplicações tecnólogicas como a computação quântica \cite{chang2019quantum}, simulação quântica \cite{georgescu2014quantum}, comunicação quântica \cite{gisin2007quantum} e metrologia quântica \cite{degen2017quantum}, que prometem moldar o século 21.

%%%%%%%%%%%%%%%%%%%%%%%%%%%%%%%%%%%%%%%%%%%%%%%%%%%%%%%%%%%%%%%%%%%%
Contudo, muito embora todas as promessas e apesar dos avanços experimentais impressionantes, os protótipos de computadores quânticos atuais são extremamente ruidosos. Processadores quânticos baseados em circuitos supercondutores, tal como aquele usado pela Google na primeira demonstração da supremacia quântica \cite{qsupremacy}, operam com uma taxa de ruído por operação quântica fundamental de aproximadamente $0.6 \%$. Em contraste, a taxa de ruído dos computadores clássicos modernos é insignificante, com componentes típicos com uma taxa de falhas da ordem de $10^{-17}$. Enquanto para os computadores clássicos a correção de erros é irrelevante, para as máquinas quânticas ela é essencial. Caso contrário, os erros se acumularão exponencialmente rápido obliterando qualquer vantangem computacional. Na prática, isso implica uma sobrecarga enorme na qual cada qubit lógico é composto de centenas ou mesmo milhares de qubits físicos.  Um computador quântico universal tolerante a falhas que possa resolver, de forma eficiente, problemas como a fatoração de inteiros e a pesquisa em um banco de dados não estruturado requerem milhões de qubits com baixas taxas de erro e longos tempos de coerência. 

Ainda que o avanço experimental para criar tais dispositivos possa levar décadas de pesquisa, a era NISQ \cite{nisq}, do inglês \textit{Noisy Intermediate-Scale Quantum}, com até algumas centenas de qubits, tempo de coerência limitado e sem correção de erros, já é uma realidade. Algo que tem motivado uma variedade de novos algoritmos para os quais pode-se garantir o desempenho na presença de erros e assim atingir vantangens computacionais \cite{qf}. Uma das principais aplicações vislumbradas para os dispositivos quânticos da era NISQ ocorre em problemas de otimização. Algoritmos tais como o QAOA \cite{qaoa}, do inglês \textit{Quantum Approximate Optimzation Algorithm}, o VQE \cite{Peruzzo_2014, McClean_2016}, do inglês \textit{Variational Quantum Eigensolver} ou aqueles baseados no anelamento quântico \cite{born1928beweis, kato1950adiabatic, Farhi00} formam a base para aplicações que variam desde simulações moleculares \cite{o2016scalable} e de novos materiais \cite{bauer2020quantum} até problemas de aprendizagem de máquina \cite{biamonte2017quantum} e análise de redes complexas \cite{paparo2013quantum}.

Algoritmos quânticos de otimização tem tido particular destaque em problemas financeiros \cite{qf2,Orus_2019}, incluindo métodos de Monte Carlo, otimização de portfólio, gestão de risco financeiro  e melhoramento de algoritmos de aprendizado de máquina \cite{Baaquie_1997,Haven2002,qf,qf2, cana16,cana17,prl122,prb100,goes2021}. Essa nova linha de pesquisa, chamada \textit{quantum finance}, tem tido um número crescente de artigos científicos \cite{Egger21, Orus_2019} e até mesmo o surgimento de startups focadas na área como, por exemplo, a DualQ \cite{DualQ} e a Multiverse \cite{multiverse}. A motivação para isso é, entre outras coisas, a promessa de resolver problemas de forma mais rápida e a necessidade de garantir a segurança de dados diante de uma nova tecnologia. 

Por ora, enquanto o desenvolvimento do hardware não é suficiente para ultrapassar a era NISQ, o desafio mais urgente é a prova de princípio de que algoritmos quânticos possam executar uma gama de tarefas com desempenho ao menos comparável ao dos computadores clássicos e entender sob quais condições vantagens quânticas podem ser esperadas. Na própria RBEF, há diversos tutoriais explicando como usar circuitos para implementar algoritmos quânticos no computador quântico da empresa \textit{IBM} \cite{Gleydson2021, Antonio2022, Rabelo2018,Jos2013,Santos2016}.
%Enfatizamos que a plataforma da \textit{IBM} é constantemente atualizada e, por isso, os artigos mais antigos estão desatualizados porém é possível ter uma ideia geral do problema.}

Dentro deste contexto, o objetivo deste tutorial é proporcionar o entendimento básico da computação quântica usando um problema central do mercado financeiro: a otimização de portfólio \cite{Markowitz1952, Rosenberg2016, cohen2020portfolio,venturelli2016job,Rebentrost2018,kerenidis2019quantum, hodson2019portfolio}. Para tanto, após uma breve introdução aos conceitos básicos em computação quântica, focaremos nossa atenção na solução baseada no QAOA \cite{qaoa, hodson2019portfolio}.
Por fim, com o intuito de melhor entender e motivar a aplicação da computação quântica para finanças, preparamos um Jupyter Notebook que pode ser acessado através da Ref. \cite{noteRBEF}, onde detalhamos conceitos centrais do mundo das finanças, além de disponibilizar códigos tanto clássicos quanto quânticos para a resolução do problema da otimização de portfólio.

\section{Computação Quântica}
\label{sec:CompQuantica}

Ao contrário da informação clássica, codificada em bits 0 ou 1, a informação quântica permite a superposição de estados, o bit quântico, também chamado de qubit \cite{schumacher1995quantum, Nielsen}, do inglês \textit{quantum bit}. O qubit é associado a qualquer sistema quântico com dois estados bem definidos, na chamada base computacional, são descritos pelos estados $\ket{0}$ e $\ket{1}$. %Tipicamente, podemos pensar no estado $|0\rangle$ com um estado menor e $|1\rangle$ ao estado de maior energia \cite{Nielsen}.
Diferentemente de bits clássicos, os qubits podem estar em uma superposição de estados, ou seja,
\begin{equation}
\ket{\psi} = a\ket{0} + b\ket{1},
\label{eq:q-bit}
\end{equation}
ou, em forma matricial,
\begin{equation}
\ket{\psi} = a \begin{bmatrix}
           1 \\
           0
         \end{bmatrix}+ b \begin{bmatrix}
           0 \\
           1
         \end{bmatrix},
\label{eq:q-bit2}
\end{equation}
onde, $a, b \in \mathbb{C}$ são amplitudes de probabilidade. Mais precisamente, se uma medição for realizada na base computacional, encontraremos o qubit no estado $\ket{0}$ com probabilidade ${|a|^2}$, ou com probabilidade ${|b|^2}=1-{|a|^2}$ para o qubit no estado $\ket{1}$. 
Note na Eq. \eqref{eq:q-bit2}, que $\ket{0} = \begin{bmatrix}
           1 \\
           0
         \end{bmatrix}$ e $\ket {1} = \begin{bmatrix}
           0 \\
           1
         \end{bmatrix}$.
Vemos, portanto, que a base computacional $\left( \ket{0},\ket{1}\right)$ é composta pelos autovetores da matriz de Pauli $Z$, apresentada adiante na Eq. \eqref{eq:portaZ}.

Em sua versão mais usual, a computação quântica trabalha com a evolução do estado descrevendo um sistema quântico. Para isso há basicamente dois paradigmas de tecnologias: a computação por portas lógicas e a computação adiabática, sobre as quais faremos uma breve descrição a seguir. Para alternativas ao modelo de computação quântica por circuito e a computação quântica adiabática, pode-se consultar as seguintes referências \cite{hans2003,nielsen2006,zanardi1999,kitaev2003}. 

\subsection{Modelo de circuitos da computação quântica}
\label{subsec:gates}

Um dos principais e mais utilizados modelos de computação quântica é aquele baseado em circuitos no qual um algoritmo é realizado através da aplicação sequencial de portas lógicas quânticas em um certo estado quântico de entrada. Tipicamente, um estado de referência onde os $n$ qubits a serem processados são inicializados no estado $\ket{0}^{\otimes n}$. O que um computador quântico que usa tecnologia de portas faz é simplesmente aplicar uma operação unitária $U$ neste estado de entrada, gerando um estado final $\ket{\psi}_{final}=U\ket{0}^{\otimes n}$ que codifica a resposta da computação que pode ser acessada através de medições apropriadas. Apesar de $U$ ser um operador atuando em um espaço de $n$ qubits, um dos resultados fundamentais da computação quântica nos mostra que esta operação unitária genérica pode ser arbitrariamente bem aproximada por operações unitárias atuando apenas em um único qubit, além de unitárias emaranhantes em dois qubits. Em outras palavras, com um conjunto finito de portas lógicas quânticas podemos realizar uma computação quântica universal. 

De modo geral, transformações em sistemas quânticos controlados, sem acesso a graus de liberdade externos, devem ser reversíveis e precisam preservar a probabilidade total dos resultados de uma medida em $100\%$. Por conta disso, portas lógicas devem corresponder a transformações unitárias no estado quântico sobre o qual atuam. Matematicamente, uma transformação é dita unitária quando seu operador adjunto correspondente é também sua transformação inversa, isto é
\begin{equation}
U^\dagger\,U = U\,U^\dagger = I,
\label{eq:Unitary}
\end{equation}
onde $U^\dagger$ é o operador adjunto de $U$, obtido a partir da transposta de $U$ tomando-se o complexo conjugado dos seus elementos, e $I$ é a matriz identidade. A reversibilidade é garantida pela existência de inversa, enquanto a preservação da normalização dos estados é dada pela Eq.\ \eqref{eq:Unitary} junto do fato que $(U\ket{\psi})^\dagger = \bra{\psi}U^\dagger$, o que leva a
\begin{equation}
    |\bra{\psi}U^\dagger U\ket{\psi}|^2 = |\braket{\psi}{\psi}|^2 = 1.
\end{equation}

%, com elementos $(U^\dagger)_{ij} = U^*_{ji}$

Pode-se mostrar que, a menos de uma fase global arbitrária, operações unitárias que atuam sobre um único qubit %possuem a propriedade particular de corresponderem
correspondem a operadores com a mesma forma de um operador de rotação de um sistema com spin $1/2$. %Temos então que
Dessa forma, toda unitária sobre um qubit pode ser descrita a partir de um eixo de rotação $\hat{n}$ e um ângulo $\theta$ na forma
\begin{equation}
U = e^{-i \theta \hat{n}\cdot\boldsymbol{\sigma}/2} = \cos(\theta/2) I - i \sin(\theta/2)\,\hat{n}\cdot\boldsymbol{\sigma},
\label{eq.OperadorEvo}
\end{equation}
onde $\hat{n}\cdot\boldsymbol{\sigma}$ é uma notação reduzida para o operador $n_x\,X + n_y\,Y + n_z\,Z$, sendo $X$ e $Z$ as matrizes de Pauli descritas respectivamente nas Eq. \eqref{eq:portaX} e \eqref{eq:portaZ} e $Y$ é a matriz de Pauli restante tal que 
\begin{equation}
Y = 
\begin{bmatrix}
0 & -i \\
i & 0
\end{bmatrix}.
\label{eq:portaY}
\end{equation}

Note, em particular, que as portas $X$, $Z$ e $H$, apresentadas a seguir, correspondem a rotações de um ângulo $\pi$ em torno dos eixos $\hat{x}$, $\hat{z}$ e $(\hat{x}+\hat{z})/\sqrt{2}$, respectivamente, a menos de uma fase global, que não interfere no resultado de medidas.

\begin{figure}[t!]
\centering
\includegraphics[width=0.4\textwidth]{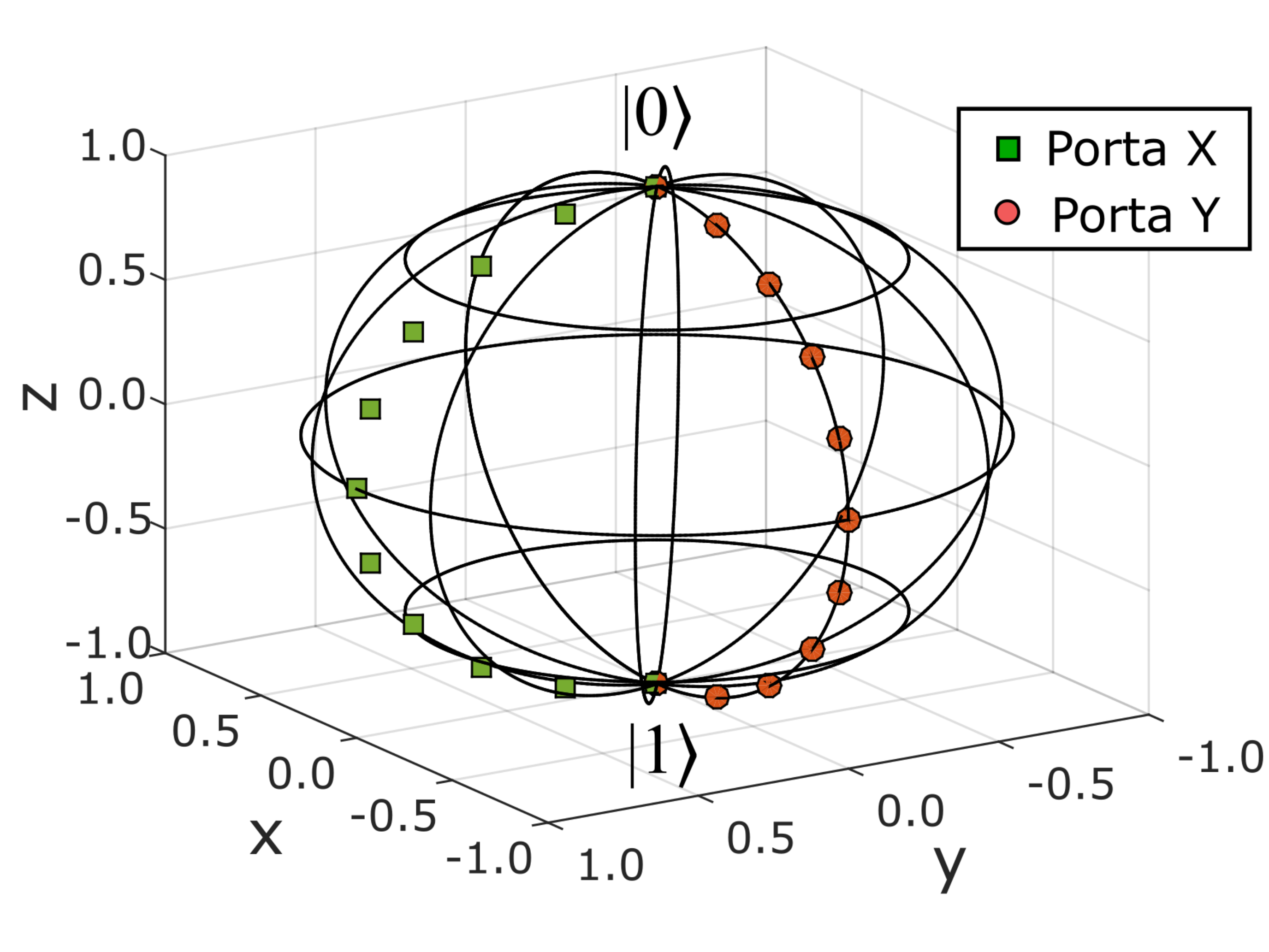}
\caption{Representação da esfera de Bloch para a implementação dos operadores $X$ e $Y$ em um qubit com estado inicial $\ket{0}$.}
\label{fig:EsferaBloch}
\end{figure}

A noção de transformações unitárias como rotações em um espaço abstrato pode ser realizada de maneira mais concreta. Cada estado de um qubit pode ser visualizado como um ponto em uma esfera, tal esfera é conhecida como esfera de Bloch.
Na Figura \ref{fig:EsferaBloch} mostramos a representação das operações das portas $X$ e $Y$ na esfera de Bloch aplicadas a partir do estado $\ket{0}$, ou seja, a trajetória da evolução de $\theta$ da Eq. \eqref{eq.OperadorEvo} quando $\boldsymbol{\sigma}$ equivale às portas $X$ (Eq. \eqref{eq:portaX}) e $Y$ (Eq. \eqref{eq:portaY}). Enfatizamos que estas evoluções ocorrem tão rapidamente que podemos considerar que a mudança de estado ocorre instantaneamente após a aplicação de $U$.

As portas apresentadas a seguir fazem parte do circuitos quânticos mostrados nas Figuras \ref{fig:circuitoQAOA} e \ref{fig:QAOA_HBeHC}, discutidos detalhadamente mais a frente quando introduzirmos o Algoritmo de Otimização Quântica Aproximada (em tradução livre do inglês \textit{Quantum Approximate Optimization Algorithm}, ou QAOA). 

As portas lógicas $X$ e $Z$, são portas de um único qubit e correspondem, respectivamente, às matrizes de Pauli $X$ e $Z$, ou seja
\begin{equation}
X = 
\begin{bmatrix}
0 & \ \ 1 \\
1 & \ \ 0
\end{bmatrix},
\label{eq:portaX}
\end{equation}
e
\begin{equation}
Z = 
\begin{bmatrix}
1 & 0 \\
0 & -1
\end{bmatrix}.
\label{eq:portaZ}
\end{equation} 

A porta $X$ é o análogo da porta clássica NOT, a porta $Z$ também é conhecida como porta de fase - por colocar um fase negativa no estado $\ket{1}$. Quando atuam na base computacional, teremos como resultado
\begin{align} 
&  X \ket{0} \rightarrow \ket{1},\\
&   X \ket{1} \rightarrow \ket{0},
\end{align}
e
\begin{align} 
&  Z \ket{0} \rightarrow \ket{0},\\ 
&  Z \ket{1} \rightarrow -\ket{1}.
\end{align}

Outra importante porta quântica é a chamada Hadamard. Representada por $H$, a porta Hadamard é uma porta de um qubit e é responsável por gerar superposições de estados a partir da base computacional $\ket{0}$ e $\ket{1}$, tal que
\begin{align} 
H \ket{0} \rightarrow \frac{1}{\sqrt{2}}\left(\ket{0} + \ket{1}\right),\\ 
H \ket{1} \rightarrow \frac{1}{\sqrt{2}}\left(\ket{0} - \ket{1}\right),
\end{align}
onde o termo $1/\sqrt{2}$ promove a normalização do estado após a aplicação da porta. A representação matricial da porta Hadamard é dada por
\begin{equation}
H = \dfrac{1}{\sqrt{2}}
\begin{bmatrix}
1 & 1 \\
1 & -1
\end{bmatrix}.
\label{Hadamard}
\end{equation}

Por fim, a que talvez seja a porta lógica quântica mais importante já que gera o emaranhamento entre os qubits, condição necessária para potenciais vantagens quântica, é a chamada CNOT. A porta Não-Controlada, ou \textit{Controlled-NOT} (CNOT), atua sobre dois qubits, um sendo chamado de qubit controle e o segundo de qubit alvo. Sua representação matricial é dada por
\begin{equation}
CNOT = 
\begin{bmatrix}
1 & 0 & 0 & 0 \\
0 & 1 & 0 & 0 \\
0 & 0 & 0 & 1 \\
0 & 0 & 1 & 0
\end{bmatrix}.
\label{eq:CNOT}
\end{equation}
Quando aplicada, por exemplo, a dois qubits no estado $\ket{10}$, teremos

\begin{eqnarray}
%\[
    CNOT\ket{10}=
    \left[\begin{array}{cccc}
    1 & 0 & 0 & 0 \\
    0 & 1 & 0 & 0 \\
    0 & 0 & 0 & 1 \\
    0 & 0 & 1 & 0
    \end{array}\right]
    \left[\begin{array}{c}
    0 \\
    0\\
    1 \\
    0
    \end{array}\right]=
    \left[\begin{array}{c}
    0 \\
    0\\
    0 \\
    1
    \end{array}\right]
%\]
\end{eqnarray}

Observe que a matriz que representa a porta CNOT (Eq. \ref{eq:CNOT}) só irá atuar no segundo qubit, já que os primeiros termos da diagonal principal representam uma identidade. No entanto, o primeiro qubit irá atuar como um qubit controle na operação e no caso dele estar no estado $\ket{0}$ nada acontece com o segundo qubit (qubit alvo). Caso contrário, se o controle está em $\ket{1}$, uma porta NOT quântica (a porta $X$) é aplicada ao qubit alvo, atuando na base computacional como
\begin{align*} 
CNOT \ket{00} \rightarrow \ket{00},\\ 
CNOT \ket{01} \rightarrow \ket{01},\\ 
CNOT \ket{10} \rightarrow \ket{11},\\ 
CNOT \ket{11} \rightarrow \ket{10}.
\end{align*}

Operações lógicas e algoritmos podem ser descritos por um diagrama similar ao usado em computação clássica, denominado circuito quântico. Em particular, na Seção \ref{sec:QAOA} explicamos mais detalhadamente sobre os elementos que compõem o circuito para implementação do algoritmo QAOA, representado nas Figuras \ref{fig:circuitoQAOA} e \ref{fig:QAOA_HBeHC}.

\subsection{Computação Quântica Adiabática}
\label{sec:QC_adia}

A computação quântica adiabática \cite{albash2018adiabatic} é um paradigma alternativo, no qual o estado fundamental de um Hamiltoniano codifica a resposta de uma dada computação. Pelo teorema adiabático, temos a garantia de que um sistema no estado fundamental de um Hamiltoniano inicial, irá se manter neste mesmo estado caso a evolução até um Hamiltoniano final seja vagarosa o suficiente, tempo este que é inversamete proporcional ao \textit{gap} entre os níveis de energia do sistema quântico em questão \cite{born1928beweis,Farhi00, adiabaticGrover,albash2018adiabatic}.

Assim, diferentemente da computação por portas, a computação adiabática tem como fundamento a evolução Hamiltoniana dependente do tempo, tal que
\begin{equation}
	H(s) = A(s)H_i + B(s)H_f,
\label{eq:Hamiltonianogeral}
\end{equation}
onde $H_i$ e $H_f$ são, respectivamente, os Hamiltonianos inicial e final. $s$ é a parametrização do tempo e varia entre 0 e 1 enquanto $A(s)$ e $B(s)$ são as funções que descrevem variação temporal do Hamiltoniano, obedecendo condições de contorno dadas por $A(0)\neq 0$, $B(1)\neq 0$, $A(1)=B(0)=0$. Ou seja, quando $t=0$ o Hamiltoniano $H_i$ prevalecerá enquanto que para $t=t_f$ o termo $H_f$ será dominante.

Segundo o Teorema Adiabático \cite{born1928beweis, kato1950adiabatic}: \textit{Um sistema físico permanece em seu autoestado instantâneo, não degenerado, ao longo da evolução temporal, se sobre ele é aplicada uma pertubação suficientemente pequena}. Portanto, para ser adiabática, a evolução do Hamiltoniano descrita na Eq.\ \eqref{eq:Hamiltonianogeral} deve acontecer da forma mais lenta possível, evitando-se transição entre níveis energéticos e garantido que o sistema permaneça em seu estado fundamental.

Pelo enunciado do Teorema Adiabático podemos entender porque tecnologia chamada \textit{annealing} quântico é comumente usada no modelo de computação adiabática. Em analogia com o termo \textit{annealing}, que em português pode ser traduzido como recozimento e, no mundo clássico, envolve problemas termodinâmicos onde um sistema é iniciado em um estado aleatório e em seguida a temperatura é reduzida de forma suficientemente lenta de modo que sempre seja garantido que o estado de maior entropia seja atingido. Durante esse processo o sistema evolui gradualmente até atingir seu estado fundamental~\cite{kirkpatrick1983optimization}.
Assim, o \textit{annealing} quântico fornece uma ponte conceitual entre a computação quântica adiabática e a otimização clássica. Contudo, na computação quântica adiabática, o termo \textit{annealing} quântico está relacionado à transição lenta entre os Hamiltonianos, portanto, neste caso, o \textit{annealing} está relacionado à mudança temporal, e não térmica, como no caso clássico.

A computação quântica adiabática é capaz de  implementar qualquer algoritmo quântico~\cite{aharonov2008adiabatic}, o que significa que também é um computador quântico universal.
Entretanto, na prática, a tecnologia de \textit{annealing} quântico nos restringe a problemas que podem ser escritos como um problema tipo Ising ou QUBO (Otimização Binária Quadrática Irrestrita, em tradução livre). 
Vale ressaltar que, com o objetivo de tornar a computação quântica adiabática mais popular, a empresa D-Wave, líder nesta área, disponibiliza uma ferramenta que fornece interoperabilidade entre o computadores quânticos que funcionam usando a técnica de \textit{annealing} e aqueles usando circuitos quânticos~\cite{DWAVE_cross_paradigm}. Mais detalhes sobre computação quântica adiabática podem ser encontrados em \cite{Souza2021}.

Como discutiremos mais abaixo, é interessante notar que o QAOA (Quantum Approximate Optimization Algorithm) é um algoritmo implementado através da computação de portas que de fato pode ser entendido como uma discretização da evolução temporal contínua de uma computação quântica adiabática.

\section{Fundamentos Básicos do Mercado Financeiro}

Dois dos elementos básicos do sistema financeiro são: (\emph{i}) Ações, que são títulos de posse de frações de uma determinada empresa e (\emph{ii}) \textit{commodities}, que são bens fungíveis, como por exemplo, ouro, açúcar e dinheiro. Nos referimos de forma genérica a esses elementos como ativos. Ambos são atribuídos de um valor relativo que varia de acordo com a oferta e procura: um produto escasso e muito demandado torna-se valioso, ao passo que outro, ofertado em quantidades que superam sua procura perde seu valor. Uma carteira, ou portfólio, de um investidor é basicamente o seu investimento em diferentes tipos de ativos. Por exemplo, se você tem investimentos em ações de 5 empresas (por exemplo: Google, Amazon, Tesla, Petrobras e Vale), as ações dessas empresas compõem sua carteira de investimentos. 

As ações, que serão os ativos de exemplo neste artigo, são, como já dito, frações do capital social de uma empresa. Em geral, as empresas oferecem tais ações aos investidores no mercado financeiro com o intuito de arrecadar recurso para expandir seu negócio. A bolsa de valores corresponde a um ambiente de negociação onde investidores podem comprar ou vender ações e diversos outros produtos financeiros como \textit{commodities} e derivativos. 

Nesta seção discutiremos os elementos básicos para entender o mundo das finanças. No material suplementar a este artigo, um Jupyter Notebook \cite{noteRBEF}, pode-se encontrar uma descrição mais detalhada de conceitos do mercado financeiro, como bolsa de valores, ações, fundos de investimentos, criptomoedas, entre outros.

\subsection{Abordagem Clássica}

Os retornos esperados de um ativo são simplesmente a média da variação percentual nos preços de suas ações. Portanto, o valor do retorno esperado que obtemos aqui são os retornos esperados diários. 
Considerando o ativo $i$, o seu retorno no dia $t$ será dado por $R_{t}^{(i)}=(P_{t+1}^{(i)}-P_t^{(i)})/P_t^{(i)}$, onde $P_t^{(i)}$ e $P_{t+1}^{(i)}$ são os preços da ação $i$ nos dias $t$ e $t+1$ \cite{Wilmott2007}. Podemos definir $R^{(i)}$ como um vetor (coluna) de retornos referente a cada tempo (dia) $t$ da ação $i$. Ou seja, cada componente do vetor $R^{(i)}$ será um valor dado por $R_t^{(i)}$. Chamaremos de $\Delta t$ o intervalo de tempo (cada dia), $t$ o tempo decorrido a partir do dia inicial, $M$ é o número total de passos (dias) considerado e $T_{tot}$ é o tempo total tal que $T_{tot}=M \Delta t$.

O retorno esperado (média dos retornos) para a ação $i$ será dado por:
\begin{equation}
\mu^{(i)} = \frac{1}{M}\sum \limits_{t=1} R_t^{(i)},
\label{eq:mu}
\end{equation}
no nosso exemplo usamos a média de retornos diários no período de um ano, ver Ref. \cite{noteRBEF} para detalhes práticos.
% No nosso exemplo usamos a média anual de retorno, ver Ref. \cite{noteRBEF} para detalhes práticos.
 
Precisamos entender como o comportamento de ações distintas estão relacionadas, para isso é preciso encontrar a matriz de covariância, esta é usada para calcular o risco de uma carteira de ações.
Podemos escrever uma matriz onde a diferença entre o retorno de cada ativo e o seu retorno médio está representado em uma coluna, ou seja
\begin{equation}
R =\left[\begin{array}{cccc}
R_{1}^{(1)}-\mu^{(1)} & R_{1}^{(2)}-\mu^{(2)} & \cdots & R_{1}^{(q)}-\mu^{(q)}\\
R_{2}^{(1)}-\mu^{(1)} & R_{2}^{(2)}-\mu^{(2)} & \cdots & R_{2}^{(q)}-\mu^{(q)}\\
\vdots & \vdots & \ddots & \vdots\\
R_{T_{tot}}^{(1)}-\mu^{(1)} & R_{T_{tot}}^{(2)}-\mu^{(2)} & \cdots & R_{T_{tot}}^{(q)}-\mu^{(q)}
\end{array}\right]_{M \times q}
\label{eq:retorno_preco}
\end{equation}
onde $q$ é número total de ativos.

A matriz de covariância será dada por
\begin{equation}
    \Sigma = \frac{1}{M-1} R^T R,
\label{eq:sigma}
\end{equation}
onde $R^T$ é a matriz transposta de $R$. $\Sigma$ é a matriz de covariância de ordem $q \times q$ onde cada uma de suas componentes será a relação entre dois ativos $i$ e $j$ tal que
\begin{equation}
    \Sigma_{ij}=\frac{1}{M-1}\sum_{t=1}^M \left[ \left(R_{t}^{(i)}-\mu^{(i)}\right)\left(R_{t}^{(j)}-\mu^{(j)}\right)\right].
\label{cov_ij}
\end{equation}
Uma covariância positiva significa que os retornos dos dois ativos se movem juntos (ambos caem ou sobem), enquanto uma covariância negativa significa que eles se movem inversamente (quando um sobe ou outro tende a cair e vice-versa). %O risco (assumido como a volatilidade) pode ser reduzido em uma carteira combinando ativos que tem uma covariância negativa.

No mercado financeiro, o risco quantifica a possibilidade de que o retorno obtido do ativo possa ser diferente da estimativa de retorno, portanto, a medida do risco depende da distribuição dos retornos. É preciso analisar o comportamento do ativo e assim estimar o risco de um portfólio de investimentos (com a combinação apropriada de ativos), algo que pode ser feito através do histórico de informações de mercado. De fato, o desvio padrão dos retornos nos fornece o grau de variação de uma série de preços de negociação ao longo do tempo, ou seja, como os retornos podem flutuar. 

\section{O problema da otimização de portfólio}
\label{sec:AplicacaoCQnoMF}

Uma figura importante para gestão de risco no mercado financeiro é a construção de portfólios. Introduzida por Markowitz \cite{Markowitz1952}, um portfólio consiste em um conjunto de $q$ ativos nos quais se deseja investir um determinado capital. A decisão de quanto e quando investir em cada ativo é tomada diante da estimativa de retorno $R^{(i)}$ e de previsão de riscos de perda $\Sigma$ \cite{Ziemann_bookFinance}. Tal processo de estimativa de risco e retorno é chamado de otimização de portfólio. Ou seja, para um determinado risco, existe um portfólio que maximiza o retorno. Por outro lado, para um determinado retorno, existe um portfólio que minimiza o risco.

O problema de otimização de portfólio pode ser definido simplesmente como um problema cuja finalidade é encontrar os valores ótimos dos pesos que maximizam os retornos esperados enquanto minimizam o risco (volatilidade).

Dado uma matriz de pesos $w$ que representará a alocação percentual dos investimentos entre essas $q$ ações, dado por \begin{equation}
w =\left[\begin{array}{cccc}
w_1 & w_2 & \cdots & w_q\end{array}\right],\label{eq:pesos}
\end{equation}
e tal que a soma deste pesos deve ser $\sum_{i=1}^q w_i=1$.
Para $q=4$ ações, por exemplo, podemos ter uma alocação uniforme $w = [0.25,  0.25, 0.25, 0.25]$ ou, inclusive, uma matriz aleatória, como mostrado na resolução prática com $30000$ portfólios, ver \cite{noteRBEF}.

O retorno esperado do portfólio $R_p$ é dado por
\begin{equation}
    R_p = w_1 \mu^{(1)} + ... w_q \mu^{(q)},
\label{eq:R_p}
\end{equation}
onde $\mu^{(i)}$ é o retorno esperado para cada ativo $i$. Observe que $R_p\equiv\mu^T w$.

A volatilidade mede a dispersão dos retornos. Ela está relacionada com a medida de risco, quanto mais o preço de uma ação varia em um determinado período de tempo, maior o risco de se ganhar ou perder dinheiro. A volatilidade do portfólio é dada por:
\begin{equation}
\sigma(R_p) = \sqrt{\sum_{i=1}^{q} \sum_{j=1}^{q} w_i w_j \Sigma_{ij}},
\label{eq:sigmaR_p}
\end{equation}
onde $w_i$ e $w_j$ denotam os pesos dos ativos de $1$ a $q$ (no nosso exemplo, de $1$ a $4$) e $\Sigma_{ij}$ é a covariância das duas ações denotadas por $i$ e $j$ (Eq. (\ref{cov_ij})).

Na prática, como existem investimentos com risco nulo, ou praticamente nulo, a carteira ideal não é escolhida apenas olhando o maior retorno ou o menor risco. De fato, há um índice denominado \textit{Sharpe Ratio} (SR) -- índice de Sharpe, em português -- que expressa o retorno médio obtido em excesso da taxa livre de risco.
A taxa de retorno livre de risco, ou fator de risco, é o retorno de um investimento com risco zero, o que significa que é o retorno que os investidores poderiam esperar sem correr risco.

Por exemplo, se um portfólio A produz um retorno $R_A = 2*R_f$ duas vezes maior que o retorno de um hipotético ativo livre de risco ($R_f$), mas com um risco $\sigma_A$ e um portfólio B um retorno $R_B = 2*R_f$ quatro vezes maior que o ativo livre de risco, mas ao custo de um risco $\sigma_B$, mas digamos que $\sigma_B > 2*\sigma_A$. Assim, a carteira B, embora com maior retorno absoluto (o dobro), não é a carteira ideal pois não ofereceu um prêmio de risco adequado, pois o risco mais que dobrou. Para mais detalhes metodológicos sofre o \textit{Sharpe Ratio}, verificar Refs. \cite{CDI,SRCDI}.

Dentro deste contexto, a carteira de risco ideal é aquela com o maior \textit{Sharpe Ratio}  \cite{Sharpe_1966}, definido como
\begin{equation}
SR = \frac{R_p - R_f}{\sigma},\label{eq:SR}
\end{equation} 
onde $R_p$ é o retorno do portfólio, $R_f$ é taxa de lucro livre de risco e $\sigma$ é o desvio padrão do portfólio (volatilidade). Cabe mencionar que retorno e risco são médias de janelas de $12$ meses, considerado o ano como tendo $250$ dias de negociação, conforme detalhado no material suplementar \cite{noteRBEF}. Outra observação importante é que, na prática, investimentos livre de risco não existem e as taxas apresentadas adiante, como CDI e outras, são meras referências numéricas com taxas semelhantes às taxas hipotéticas.

Um dos pontos fortes do trabalho de Markowitz foi mostrar que a busca por taxas de retorno maiores implica necessariamente em montar portfólios com maiores níveis de risco. Em termos práticos, se buscamos um maior retorno, precisaremos expor nossa carteira de investimentos a um risco mais elevado, que significa maiores variações percentuais, ou seja, uma maior volatilidade.

\subsection{Exemplo prático}

De um modo geral, um bom portfólio é aquele que nos fornece o máximo retorno com um mínimo de risco. Assim, assume-se que os investidores são avessos ao risco, ou seja, se houver uma escolha entre carteiras de baixo e alto riscos com os mesmos retornos, um investidor escolherá uma com baixo risco. Portanto, para decidirmos, entre a infinidade de combinações possíveis de carteiras, aquela que é ótima, i.e. para fazer a otimização de portfólio, devemos escolher a carteira com máximo retorno dado um certo grau de risco. 

\begin{figure}[t!]
\centering
\includegraphics[width=0.475\textwidth]{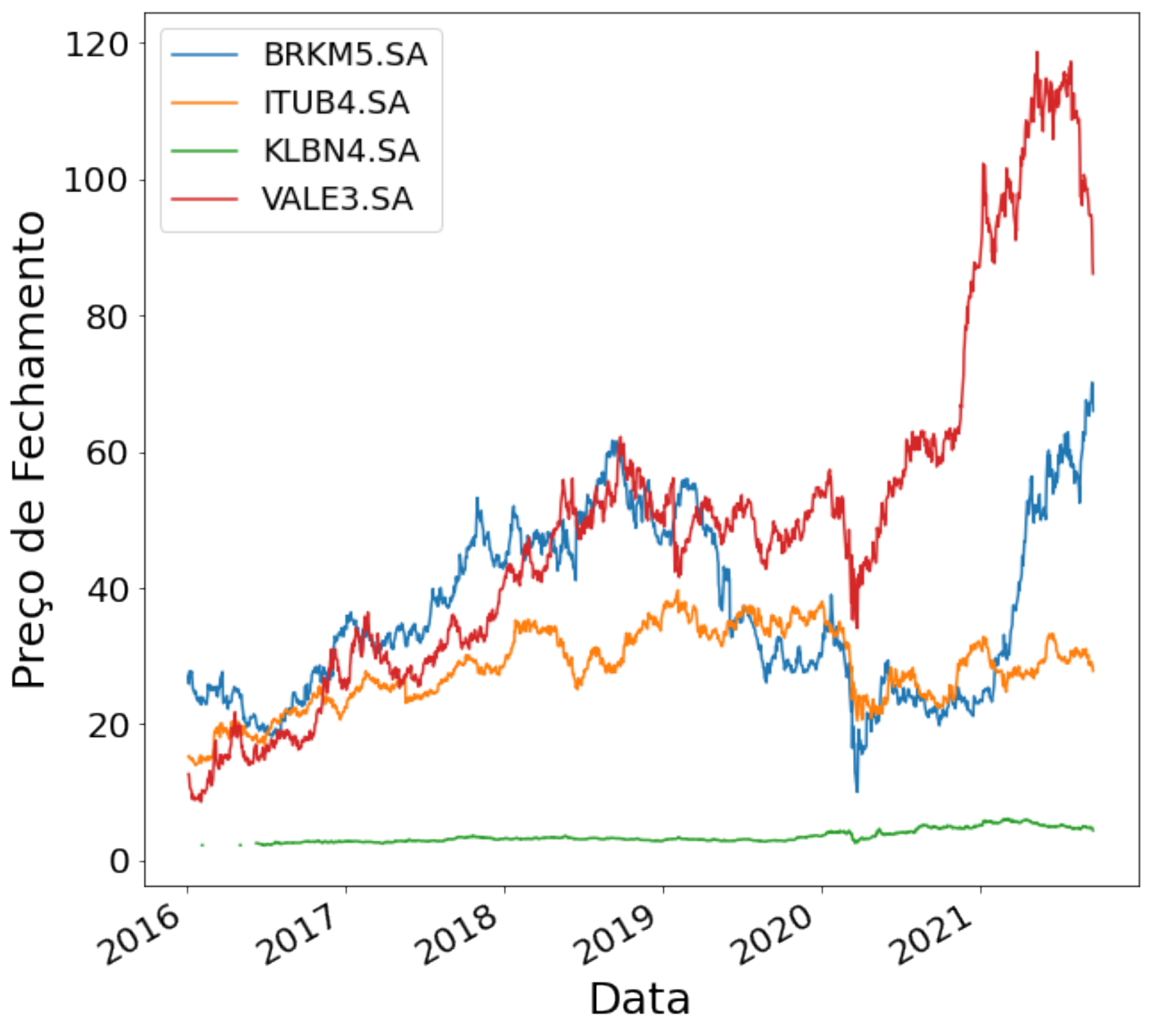}
\caption{Exemplo da evolução temporal do preço de fechamento de ativos (disponíveis na bolsa brasileira) que formam uma carteira composta com ações das empresas Braskem, Itaú, Klabin e Vale.}
\label{fig:prices}
\end{figure}

Como exemplo, consideramos  os preços diários das ações das empresas Braskem (BRKM5), Itaú (ITUB4), Klabin (KLBN4) e Vale (VALE3)  para o período entre 01 de janeiro de 2016 até 20 de setembro de 2021 obtidos com o pacote ``yfinance'' provido pelo \textit{Yahoo! Finance}, cujos preços estão apresentados na Fig. \ref{fig:prices}.

Conforme discutido, precisamos obter dois ingredientes fundamentais: o vetor de retorno para cada ativo e a matriz de covariância para uma dada periodicidade (ver Ref. \cite{noteRBEF} para mais detalhes). Para o intervalo estabelecido e a periodicidade anual, o vetor de retorno médio anual e a matriz de covariância são apresentados, respectivamente, nas Tabelas \ref{tab:retornos} e \ref{tab:matriz_cov} conforme calculado na Ref. \cite{noteRBEF}.

\begin{table}[t]
\begin{centering}
\begin{tabular}{|c|c|c|c|c|}
\hline 
 Ativos & BRKM5 & ITUB4 & KLBN4 & VALE3 \tabularnewline
\hline 
\hline 
Retorno ($\mu$) & 0.32 & 0.06 & 0.11 & 0.30\tabularnewline
\hline 
\end{tabular}
\par\end{centering}
\caption{Retorno anual médio dos preços de fechamento das ações da Braskem (BRKM5), Itaú (ITUB4), Vale (VALE3) e Klabin (KLBN4) referentes ao período entre 01 de janeiro de 2016 até 20 de setembro de 2021.}
\label{tab:retornos}
\end{table}

Usando a teoria moderna de portfólio, conhecida também como a abordagem de Markowitz, nós geramos $30000$ vetores pesos $w \in \mathbb{R}^4$ aleatoriamente, onde cada componente de um vetor específico representam, nesta ordem, a alocação em Braskem, Itaú, Klabin e Vale. Em cada iteração, consideramos os diferentes pesos individuais para cada ativo e calculamos o retorno e a volatilidade dessa alocação particular do portfólio. Assim, para cada $w$, calculamos o retorno anual da carteira por meio da Eq. \eqref{eq:R_p} e dos retornos individuais apresentados na Tabela \ref{tab:retornos}, bem como o risco (volatilidade) anual atráves da Eq. \eqref{eq:sigmaR_p} e dos dados da Tabela \ref{tab:matriz_cov}. Ver Ref. \cite{noteRBEF} para maiores detalhes práticos de todos os exemplos dessa seção.

\begin{table}[h]
\begin{centering}
\begin{tabular}{|c|c|c|c|c|}
\hline 
 Ativos & BRKM5 & ITUB4 & KLBN4 & VALE3 \tabularnewline
\hline 
\hline 
BRKM5 & 1.00 & 0.37 & 0.28 & 0.33\tabularnewline
\hline 
ITUB4 & 0.37 & 1.00 & 0.18 & 0.36\tabularnewline
\hline 
KLBN4 & 0.28 & 0.18 & 1.00 & 0.24\tabularnewline
\hline 
VALE3 & 0.33 & 0.36 & 0.24 & 1.00\tabularnewline
\hline 
\end{tabular}
\par\end{centering}
\caption{Matriz de covariância dos preços de fechamento das ações da Braskem (BRKM5), Itaú (ITUB4), Vale (VALE3) e Klabin (KLBN4) referentes ao período entre 01 de janeiro de 2016 até 20 de setembro de 2021.}
\label{tab:matriz_cov}
\end{table}

\begin{table*}[t]
\begin{centering}
\begin{tabular}{|c|c|c|c|c|c|c|}
\hline 
 Descrição do Portfólio & BRKM5 & ITUB4 & KLBN4 & VALE3 & Risco & Retorno \tabularnewline
\hline 
\hline 
Risco Mínimo & 0.009 & \textbf{0.404} & \textbf{0.503} & 0.083 & 0.244 & 0.108 \tabularnewline
\hline 
Retorno Máximo & \textbf{0.919} & 0.006 & 0.009 & \textbf{0.066} & 0.492 & 0.311 \tabularnewline
\hline 
Melhor $SR$ $(R_f = 0.015)$ & \textbf{0.370} & 0.000 & 0.193 & \textbf{0.436} & 0.347 & 0.270\tabularnewline
\hline 
Melhor $SR$ $(R_f = 0.028)$ & \textbf{0.403} & 0.002 & 0.099 & \textbf{0.496} & 0.372 & 0.287 \tabularnewline
\hline 
Melhor $SR$ $(R_f = 0.100)$ & \textbf{0.451} & 0.000 & 0.022 & \textbf{0.526} & 0.395 & 0.303 \tabularnewline
\hline
\end{tabular}
\par\end{centering}
\caption{Apresentação de alguns portfólios conforme destacados na Fig. \ref{fig:fronteira}, contendo: \textit{i}) as frações (ou pesos) de cada ação na carteira, bem como \textit{ii}) o risco e \textit{iii}) o retorno correspondentes, obtidos usando a abordagem clássica de Markowitz. Em destaque, as duas maiores alocações de cada carteira.}
\label{tab:port}
\end{table*}

\begin{figure}[t!]
\centering
\includegraphics[width=0.475\textwidth]{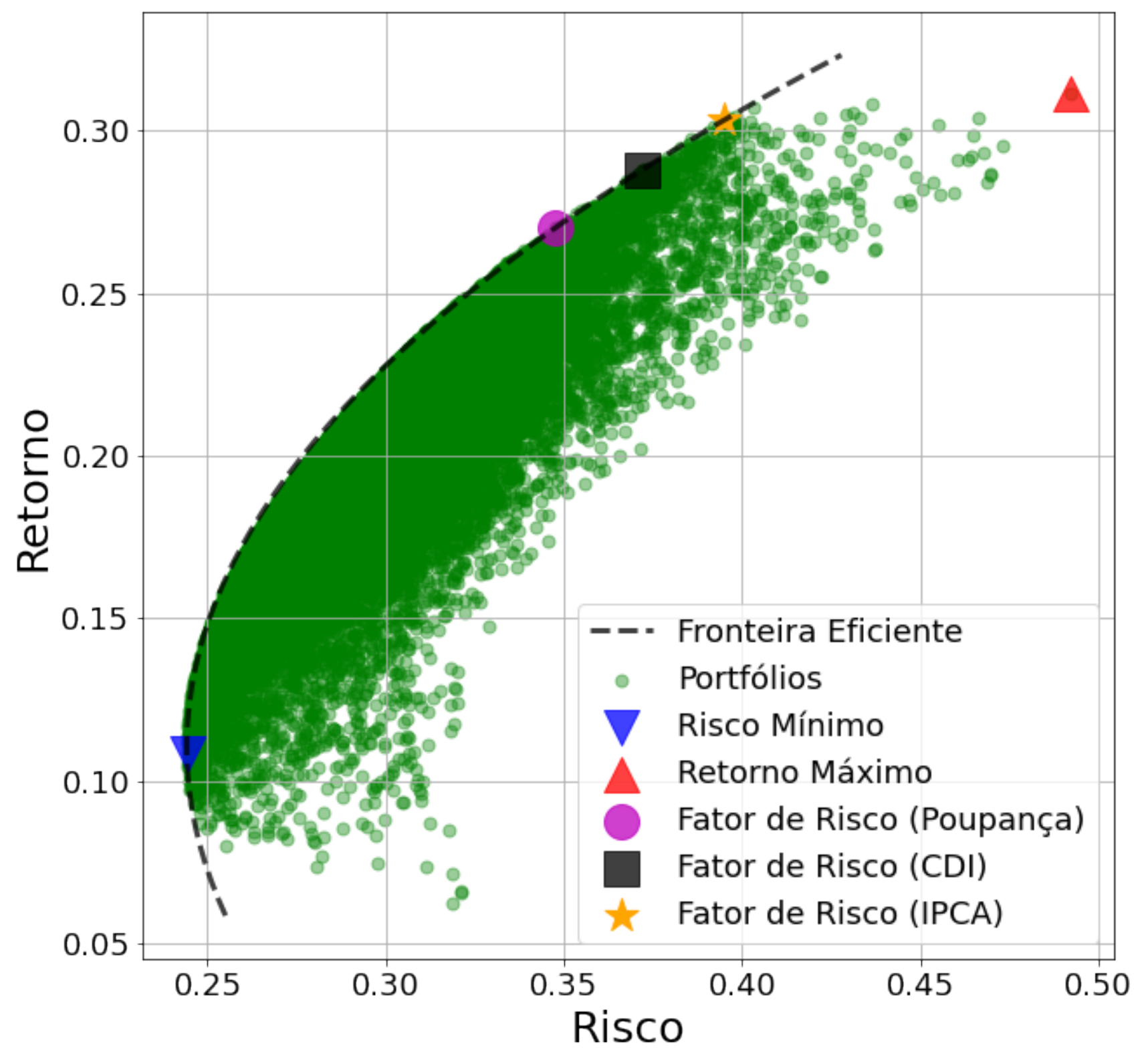}
\caption{O conflito entre Risco ($\Sigma$) e Retorno ($\mu$) ilustrado com o gráfico da Fronteira Eficiente. Cada ponto verde representa uma carteira com diferente proporções (pesos) de ações das empresas Braskem, Itaú, Klabin e Vale. Observe como a busca por retornos maiores implica em assumir maiores níveis de risco. Confira a Ref. \cite{noteRBEF} para verificar a obtenção da Fronteira Eficiente na prática.}
\label{fig:fronteira}
\end{figure}

Observe que a criação da carteira, nesta vertente clássica, consiste em determinar o percentual que aplicamos em cada uma das ações (BRKM5, ITUB4, KLBN4 e VALE3). Através do plot do Retorno em função do Risco para os $30000$ portfólios (ver Fig. \ref{fig:fronteira}), podemos determinar carteiras com diversas características. Podemos, por exemplo, selecionar a carteira com retorno máximo para um determinado risco ou uma carteira com risco mínimo para um determinado retorno ou ainda, podemos simplesmente selecionar a carteira com Sharpe Ratio máximo para determinados fatores de risco, como definido na Eq. \eqref{eq:SR}. 

Tais portfólios ideais para distintos níveis de risco formam o que se chama de curva de Fronteira Eficiente, representada na linha tracejada na Fig. \ref{fig:fronteira}. Sobre tal curva encontram-se os portfólios com a melhor relação risco/retorno, sendo mais vantajosos que todos os demais portfólios situados abaixo dela. Como precisaremos de alguns pontos (portfólios) para traçar a curva de Fronteira Eficiente, vamos explicar alguns deles. Na Tabela \ref{tab:port} é feita uma compilação de alguns portfólios, conforme destacados na Fig. \ref{fig:fronteira}, com as frações (ou pesos) de cada ação na carteira, bem como o risco e o retorno correspondentes.

Após o cômputo do risco e do retorno para cada um dos $30000$ pesos gerados aleatoriamente, podemos investigar a composição de cada um dos portfólios. Como sabemos, um portfólio de interesse é o com menor risco que, na Figura \ref{fig:fronteira}, corresponde ao triângulo para baixo em azul. Na Tabela \ref{tab:port}, mostra-se que equivale a um risco de cerca de $24.4 \%$, significando que a carteira variou em média em torno de $24.4 \%$ por ano, com um retorno anual médio de cerca $10.8 \%$. A composição desta carteira é concentrada em Itaú ($40.4\%$) e Klabin ($50.3 \%$).

Outro portfólio natural a ser investigado é a carteira com retorno máximo, que corresponde ao triângulo para cima em vermelho na Fig. \ref{fig:fronteira}. Vemos, na Tabela \ref{tab:port}, que a composição deste portfólio é fortemente concentrado em Braskem ($92 \%$), com uma alocação residual em Vale ($6.6 \%$), o que já era esperado dado o destaque em perfomance das ações da Braskem e Vale com relação às demais ações apresentadas na Fig. \ref{fig:prices}. A carteira apresenta retorno anual de cerca de $31 \%$ com risco de $50 \%$. Observe que o risco é aproximadamente o dobro da carteira de risco mínimo, porém com o triplo do retorno.

Seguindo a hipótese de que os investidores são naturalmente avessos ao risco, surge a necessidade de comparar o desempenho dos portfólio com investimentos livres de risco, através do cálculo do Sharpe Ratio (SR), conforme já prescrito na Eq. \eqref{eq:SR}. Suponha que exista no mercado uma Taxa Livre de Risco equiparada com a taxa de poupança, CDI e IPCA. Na Figura \ref{fig:fronteira}, para descobrir o portfólio com máximo SR, usamos três fatores de risco $R_f$ (três cenários qualitativamente distintos): a taxa de poupança, a CDI e o IPCA, aproximadamente. Note que, para cada fator de risco, temos um porfólio ideal, que não coincide com o portfólio de retorno máximo. Ou seja, para bater um benchmark cada vez mais alto (taxa de retorno sem risco crescente), precisamos tomar riscos cada vez maiores, caso o apetite do investidor não seja satisfeito com a taxa livre de risco.

Em um cenário onde a taxa livre de risco disponível seja em torno de $R_f = 1.6\%$, como a nossa poupança atual, o portfólio com maior SR, após testar $30000$ composições (pesos) de portfólios, de igual modo, com Braskem, Itaú, Klabin e Vale apresenta a seguinte composição para proporção de alocação $w$, nesta ordem, $w = [37\%, 0\%, 19\%, 43\%]$, com um risco de cerca de $35 \%$ e um retorno de $ 27 \%$. Veja o círculo roxo na Fig. \ref{fig:fronteira} e a terceira linha de dados na Tab. \ref{tab:port}. Note que mais de $80 \%$ da alocação do recursos concentra-se em Braskem e Vale.

Para um fator de risco de $R_f = 2.86\%$ (aproximadamente a CDI), obtemos os seguintes pesos de alocação correspondente a um SR máximo: $w = [40.3\%, 0\%, 10\%, 49.7\%]$. Tal portfólio é indicado na Fig. \ref{fig:fronteira} pelo quadrado preto. Observe que quase $90\%$ da carteira é concentrada em Braskem e Vale. O retorno do portfólio é de $28.7\%$ e o risco gira em torno de $37\%$.

Por fim, para um cenário onde exista investimento livre de risco com taxa $R_f = 10\%$ (e.g., aproximadamente o IPCA), temos $w = [45.1\%, 0\%, 2.2\%, 52.6\%]$, representado pela estrela laranja na Fig. \ref{fig:fronteira}. Em primeiro lugar, observe que quase $98 \%$ da carteira é alocada, de igual modo, em Braskem e Vale. O retorno é praticamente igual ao da carteira com retorno máximo, cerca de $30.3 \%$, porém o com um risco cerca $25 \%$ menor, em torno de $40 \%$.

Apresentados tais pontos (portfólios) contidos na Fronteira Eficiente, podemos proceder um ajuste polinomial para a determinação precisa da curva. Confira nosso Jupyter Notebook \cite{noteRBEF} onde apresentamos um ajuste polinomial de grau $2$ do Risco em função do Retorno para obtenção da Fronteira Eficiente. 

A dificuldade computacional que envolve o problema de otimização de portfólio aumenta quando se adicionam restrições (como o da Eq. \eqref{eq:restricaoPortfolio}) mais realistas ao problema. Por exemplo, pode-se desejar impor restrições de positividade (de modo que só se possa comprar e não vender ativos), ou restrições de número inteiro (de modo que só se pode investir em ativos em incrementos fixos), ou limites superiores sobre o valor que se pode investir em um determinado ativo (devido, por exemplo, à liquidez). Além disso, os retornos ou correlações entre os ativos podem variar no tempo, podendo haver custos de transação para a compra e venda de ativos. 

Com essas restrições mais realistas, o problema de otimização de portfólio pode se tornar substancialmente mais desafiador de resolver, mesmo aproximadamente, e é um desafio das finanças computacionais modernas.
Uma variedade de abordagens algorítmicas quânticas \cite{nisq_algo,bouland2020prospects,moll2018,Rebentrost2018,egger2019credit,Woerner2019,stampa2020,Zoufal2019} foram desenvolvidas para resolver, ao menos aproximadamente, este problema. Abaixo definimos  matematicamente o problema de otimização, para o qual analisaremos a solução quântica proporcionada pelo QAOA.

Considere o seguinte problema de otimização de portfólio, cujo objetivo é minimizar a expressão
\begin{equation}
    \min_{x \in \{0, 1\}^n} b x^T \sigma x - \mu^T x
\label{eq:OtimizacaoRestrita}
 \end{equation}
com a restrição 
 \begin{equation}
     1^T x = B,
\label{eq:restricaoPortfolio}
 \end{equation} 
onde $x \in \{0,1\}^n$ é um vetor de variáveis binárias no qual quando $x = 1$ indica quais ativos escolher e $x = 0$ indica quais não escolher. O parâmetro $b>0$ é um controle de risco do gestor de ativos, já que ao aumentar  $b$ precisamos de maiores retornos para alcançar o mínimo da função objetiva. Fato que, pela fronteira eficiente discutida anteriormente (Fig. \ref{fig:fronteira}), exige maiores riscos. O parâmetro $B$ denota o orçamento, neste caso específico, o número de ativos a serem selecionados dentre $q$ disponíveis. Note que, com variáveis binárias, a otimização apenas nos diz quais ativos devemos investir, mas não a quantidade ou proporção com a qual devemos fazê-lo. 

 A restrição estabelecida na Eq. \eqref{eq:restricaoPortfolio} pode ser entendida como oriunda de um termo de penalidade do tipo 
 \begin{equation}
    P(x)=\alpha (B - f(x))^2
 \end{equation}
 onde $f(x)$ é uma função de penalidade e $\alpha$ é um coeficiente de escala de penalidade. Quando $f(x)=B$, $P(x)=0$ então a restrição é atendida. Quando $f(x)\neq B$, $P(x)>0$ então a restrição é violada \cite{hodson2019portfolio}. No nosso exemplo da Eq. \eqref{eq:restricaoPortfolio}, $f(x)=1^T x$. Observe que o termo de regularização penaliza carteiras com um número de ativos diferente de $B$. Veja \cite{noteRBEF} para conferir na prática. Caso o leitor(a) deseje assumir uma taxa de retorno livre de risco diversa das que utilizamos como exemplo, ele(a) pode executar o código com qualquer valor que desejar, fazendo as mudanças correspondentes no código disponibilizado. Para uma lista de títulos públicos mundiais e o os respectivos rendimentos, sugerimos consultar Ref. \cite{bondsworld}.
 
\section{O algoritmo QAOA}
\label{sec:QAOA}

O Algoritmo de Otimização Quântica Aproximada (em tradução livre do inglês \textit{Quantum Approximate Optimization Algorithm}, ou QAOA) é um algoritmo híbrido (isto é, parcialmente quântico e clássico) que busca encontrar soluções aproximadas para problemas de otimização, sendo bastante adequado para resolver problemas de otimização combinatória. 

Proposto originalmente por Farhi \textit{et al.} \cite{qaoa} em 2014, o algoritmo segue o modelo de um método variacional, onde o problema de otimização é mapeado em um Hamiltoniano de custo, $H_C$, que passa a codificar a solução do problema no seu menor autovalor. Busca-se, então, minimizar o valor esperado desse Hamiltoniano, $\langle \psi | H_C | \psi \rangle$, sobre estados $|\psi\rangle$ que pertençam a uma determinada família de estados parametrizados a partir da otimização (clássica) dos parâmetros, na expectativa de encontrar uma cota superior ao mínimo verdadeiro do espectro de $H_C$, ou o próprio valor mínimo, atingível apenas com o autoestado correspondente. 

O QAOA propõe uma forma específica de produzir estados quânticos parametrizados com uma garantia de eventual convergência ao valor ótimo do problema para uma quantidade suficientemente grande de parâmetros. Por sua simplicidade e possibilidade de aplicação imediata nos primeiros computadores quânticos e pelas garantias teóricas de convergência e de eficiência \cite{Egger21, nisq_algo}, um crescente interesse da comunidade em algoritmos variacionais vem ocorrendo ao longo dos últimos anos \cite{nisq_algo,bouland2020prospects,moll2018,Rebentrost2018,egger2019credit,Woerner2019,stampa2020,Zoufal2019}.

Apesar de ter aplicabilidade geral, o algoritmo do QAOA é comumente usado em problemas de otimização combinatória \cite{nisq_algo,bouland2020prospects}, isto é, problemas onde buscamos minimizar (ou maximizar) funções que tomam variáveis binárias como argumentos. A razão para isso tem a ver com a facilidade em codificar este tipo de problema no Hamiltoniano de custo correspondente e também com o fato de problemas dessa forma nem sempre possuírem algoritmos clássicos eficientes para sua solução. 

Na falta de um algoritmo clássico eficiente para a solução de um problema envolvendo $n$ variáveis binárias $x_i$, pode-se pensar em uma estratégia de varredura sobre todas as $2^n$ combinações de valores possíveis, ou seja, uma solução por ``força bruta''. Mesmo que se leve um tempo pequeno para cada teste individual, como por exemplo $1\,\mu s$, o crescimento exponencial do número de possibilidades rapidamente leva a tempos impraticáveis para o uso desse método. Enquanto um problema de $30$ variáveis, por exemplo, levaria em torno de $17$ minutos para ser resolvido nesse cenário, o mesmo problema com $50$ variáveis já levaria mais de $35$ anos. Compare isso com um algoritmo que levasse um tempo proporcional ao quadrado do número de variáveis, o tempo necessário para resolução do problema com $50$ variáveis seria apenas $(50/30)^2 \approx 2.77$ vezes mais demorado que o problema de $30$ variáveis. Supondo ainda $17$ minutos para este último problema, levaríamos agora em torno de $47$ minutos  para o problema maior, o que é consideravelmente menor que o esperado pelo método de força bruta. Apesar de não se acreditar que algoritmos quânticos possam reduzir drasticamente o tempo de \emph{todos} os problemas que tenham esse incremento exponencial a um problema com incrementos polinomiais, entende-se que em alguns casos alguma vantagem pode ser apresentada em relação a algoritmos clássicos, o que tornaria tratáveis algumas instâncias do problema.

\begin{figure}[t]
\centering
\includegraphics[width=0.475\textwidth]{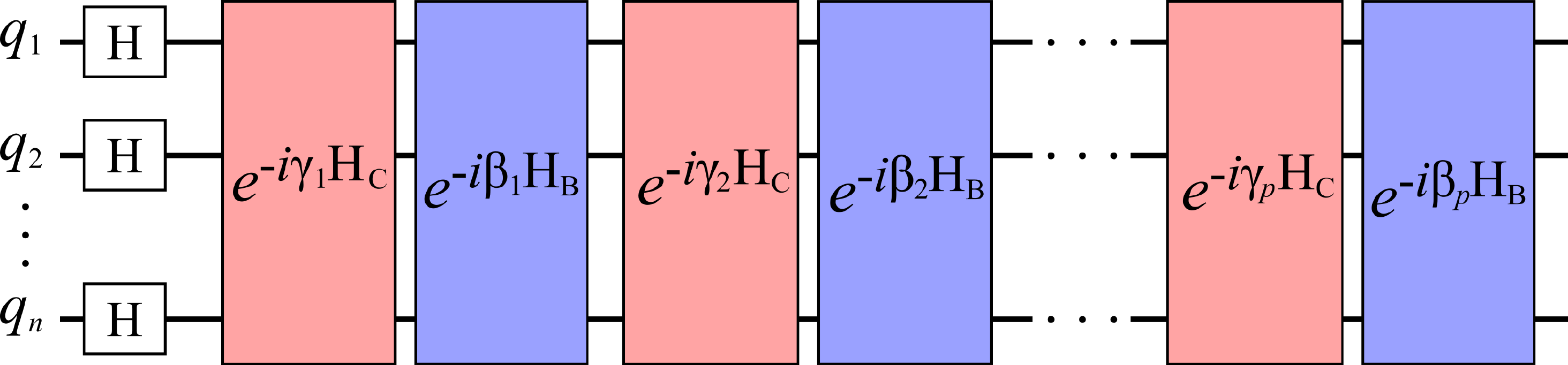}
\caption{Circuito para implementação do QAOA.}
\label{fig:circuitoQAOA}
\end{figure}

A parametrização proposta pelo QAOA se baseia fundamentalmente na aplicação sucessiva de duas operações unitárias: a unitária de custo, $U_C$, correspondente à evolução do Hamiltoniano de custo $H_C$ e a unitária de mistura, $U_B$, correspondente ao Hamiltoniano de mistura, $H_B$, dado por
\begin{equation}
H_B = \sum_{i=1}^n X_i,
\end{equation}
onde $X_i$ é o operador de Pauli $X$ (porta $X$) atuando no espaço do $i$-ésimo qubit. A ``duração'' da evolução de cada Hamiltoniano corresponde aos parâmetros a serem otimizados. Considerando parâmetros reais $\beta$ e $\gamma$, temos as operações unitárias
\begin{align}
U_C(\gamma) &\coloneqq e^{-i \gamma H_C}, \\
U_B(\beta)  &\coloneqq e^{-i \beta H_B}.
\end{align}

O algoritmo então prescreve a preparação de estados $|\psi(\boldsymbol{\gamma},\boldsymbol{\beta})\rangle$, com $\boldsymbol{\gamma} = (\gamma_1,\ldots,\gamma_p)$ e $\boldsymbol{\beta} = (\beta_1,\ldots,\beta_p)$, a partir de um estado inicial $|\psi_{\textrm{ini}}\rangle$ (também prescrito pelo algoritmo), de acordo com a sequência de evoluções unitárias
\begin{equation}
|\psi(\boldsymbol{\beta},\boldsymbol{\gamma})\rangle = U_B(\beta_p) U_C(\gamma_p) \ldots U_B(\beta_1) U_C(\gamma_1)|\psi_{\textrm{ini}}\rangle,
\label{eq:EstadoFinal_QAOA}
\end{equation}
e então buscamos encontrar o valor da expressão
\begin{equation}
F_p \coloneqq \min_{\boldsymbol{\beta},\boldsymbol{\gamma}}\, \langle \psi(\boldsymbol{\beta},\boldsymbol{\gamma}) | H_C | \psi(\boldsymbol{\beta},\boldsymbol{\gamma})\rangle,
\label{eq:OtimoQAOA}
\end{equation}
a partir de algum algoritmo clássico de otimização sobre os parâmetros. Vale notar que, como o espectro de $H_B = \sum_i X_i$ contém apenas valores inteiros (mais precisamente, $\{-1,\,1\}$), então os valores de $\beta$ podem ser tomados no intervalo $[0,\,\pi]$. Os valores de $\gamma$ podem ser buscados no intervalo $[0,\,2\pi]$.

%{\color{red}[Sugiro transferir o trecho em azul anterior para cá. Me parece mais pertinente ilustrar o circuito implementando $U_C$ e $U_B$ após falar desses operadores. Além disso, foi usado o termo "chaves lógicas", isso é o mesmo que portas lógicas? Talvez seja melhor manter como "portas lógicas" para ficar igual ao restante do texto.]}

\begin{figure}
\centering
\includegraphics[width=0.475\textwidth]{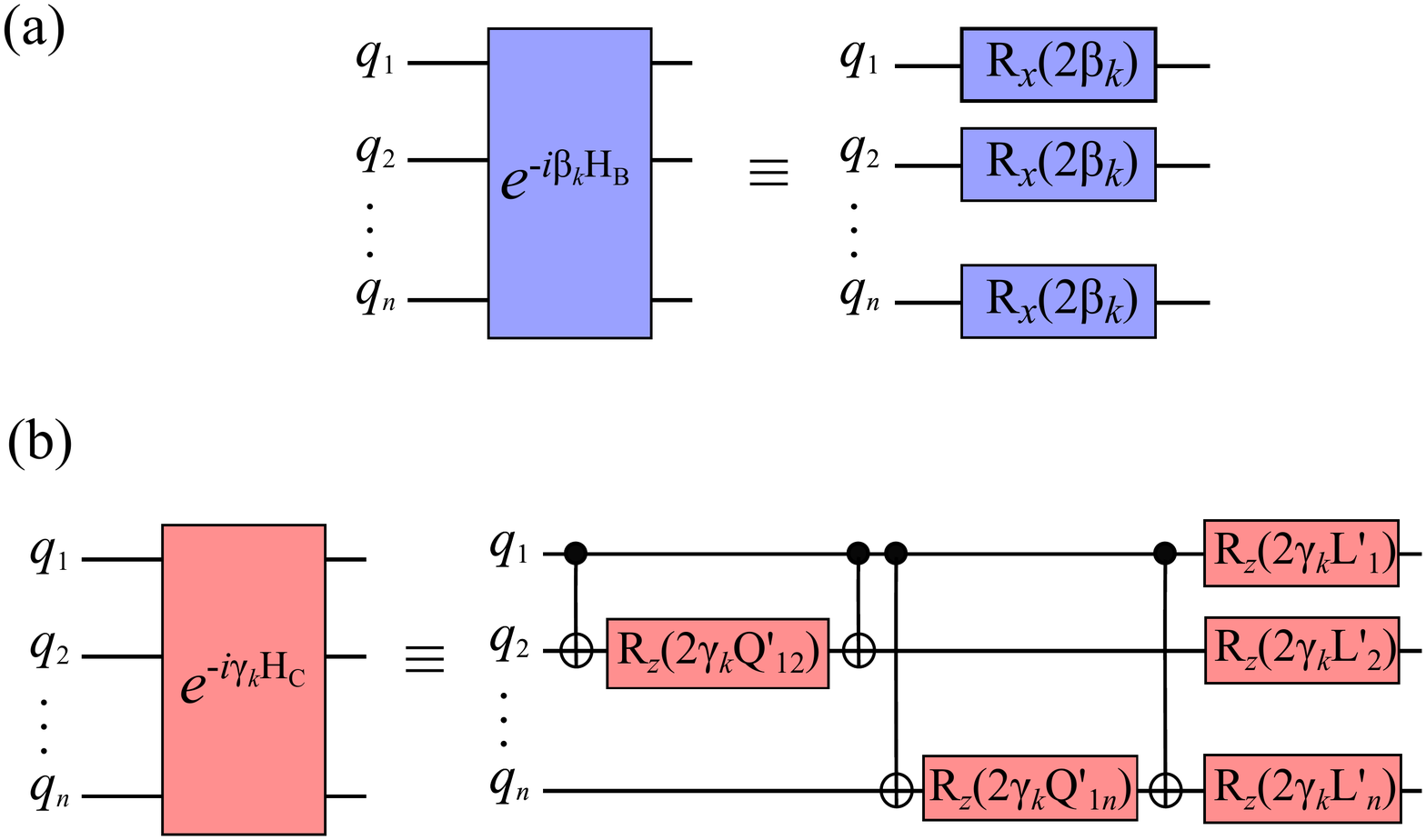}
\caption{Circuitos implementando as unitárias (a) $U_B$ e (b) $U_C$.}
%\textbf{(a)} Cada termo da Hamiltoniana de mistura atua em um qubit distinto. A unitária corresponde então a rotações em torno do eixo x para cada qubit, que podem ser executadas simultaneamente. \textbf{(b)} Usamos o modelo estabelecido para a Hamiltoniana $H_C$ na Eq. \eqref{eq:HamiltonianaQUBO}, considerando como exemplo um problema específico onde $Q'_{13}$, $L'_2$ e $K$ são nulos. Note que podemos agregar os valores de $Q'_{ij}$ e $Q'_{ji}$ em um só coeficiente, uma vez que $Z_i$ e $Z_j$ comutam entre si. Também podemos incorporar termos diagonais nos termos lineares, no problema original, uma vez que variáveis binárias satisfazem $x_i^2 = x_i$. Usamos também a identidade \eqref{eq:ident_ZZ_CNOT} para implementar as unitárias associadas aos termos quadráticos
\label{fig:QAOA_HBeHC}
\end{figure}

Na Fig. \ref{fig:circuitoQAOA} apresentamos o circuito de implementação do QAOA. Um circuito quântico determina quais e em que ordem as chaves lógicas são aplicadas a um ou mais qubits de um sistema. Estes são compostos de linhas e símbolos onde as linhas representam os qubits (uma linha para cada qubit) e cada símbolo representa uma chave lógica. Por sua vez, as chaves lógicas (descritas por caixas, com letras) descrevem um conjunto de operações quânticas aplicadas a um ou mais qubits. No circuito do QAOA vemos que temos \textit{n} qubits, primeiramente aplicamos a porta Hadamart em todos os qubits. Em seguida intercalamos operações que implementam o Hamiltoniano de mistura ($U_B$) - em azul - e o Hamiltoniano de custo ($U_C$) - em vermelho. Na Fig. \ref{fig:QAOA_HBeHC} mostramos com mais detalhes as etapas para implementação de $U_B$ e $U_C$.

O algoritmo toma como estado inicial a superposição de todos os estados da base computacional com igual amplitude, isto é, 
\begin{equation}
|\psi_{\textrm{ini}}\rangle = \sum_{x_i \in \{0,1\}} |x_1 \ldots x_n \rangle/\sqrt{2^n}.
\label{eq:superposicao}
\end{equation}
Note que este estado pode ser preparado de forma imediata a partir do estado $|0\rangle^{\otimes n}$ e da porta Hadamard (Eq. \eqref{Hadamard}), aplicando-a independentemente sobre todos os qubits. 
O resultado da combinação é a superposição de todas as possibilidades de estados da base computacional para todos os qubits, todos com a mesma amplitude. Isto é representado na equação a seguir e também na parte inicial do circuito na Fig. \ref{fig:circuitoQAOA}, tal que
\begin{equation}
|\psi_{\textrm{ini}}\rangle = H^{\otimes n}|0\rangle^{\otimes n} = \left(\frac{|0\rangle + |1\rangle}{\sqrt{2}}\right)^{\otimes n}.
\end{equation}
Veremos mais adiante que há um motivo para estabelecer essa prescrição para o estado inicial, exatamente como o autoestado de maior autovalor do Hamiltoniano de mistura, $H_B$, o que está conectado com o modelo de computação adiabática.

Outro ponto interessante de notar é que o aumento no número de parâmetros usados, implicitamente estabelecido acima como $2p$ ($p$ valores de $\gamma$ e $p$ valores de $\beta$), tende a melhorar a aproximação do valor ótimo real do problema (embora na prática possam haver flutuações, como mostrado na Fig. \ref{fig:prob}): a estrutura sequencial de operadores unitários faz com que uma preparação envolvendo $p+1$ iterações de $U_C$ e $U_B$ sempre inclua preparações com $p$ iterações, uma vez que, se usarmos $\gamma_{p+1} = 0$ e $\beta_{p+1} = 0$, obtemos $U_C(\gamma_{p+1}) = I$ e $U_B(\beta_{p+1}) = I$, de modo que
\begin{equation}
F_{p+1} \leq F_p,
\end{equation}
lembrando que $F_p$ é o valor resultante da otimização sobre os parâmetros, como definido na Eq. \eqref{eq:OtimoQAOA}. Veremos que, também por conta da conexão com o modelo de computação adiabática, de fato esta melhora não só deve ser garantida para um número suficientemente grande de $p$, como deve convergir ao valor real do problema original.
 
Particularizando o QAOA para problemas de otimização combinatória, temos ainda uma maneira concreta de estabelecer o Hamiltoniano de custo como um operador diagonal na base computacional e, portanto, em termos de combinações de produtos de operadores de Pauli $Z_i$, como num modelo de Ising. Para ilustrar isso e também por ser mais pertinente a problemas de otimização de portfólio, consideramos mais particularmente problemas tipo QUBO, onde a função custo a ser minimizada pode ser descrita como
\begin{equation}
C(x_1,\ldots,x_n) = \sum_{i,j=1}^n Q_{i,j}\,x_i x_j + \sum_{i=1}^n L_i\,x_i.
\end{equation}
 
Mapeando as variáveis binárias clássicas $x_i$ em operadores quânticos na base $z$, ou seja, $\hat{x}_i = |1\rangle\langle 1| = (I - Z_i)/2$, obtemos a propriedade
\begin{equation}
\hat{x}_i |x\rangle = x |x\rangle,
\end{equation}
onde $x \in \{0,\,1\}$. Realizando esse mapa em todas as variáveis da função custo $C$ acima, obtemos o Hamiltoniano $H_C$ desejada, que possui agora a propriedade
\begin{equation}
H_C |x_1 \ldots x_n\rangle = C(x_1 \ldots x_n) |x_1 \ldots x_n\rangle,
\end{equation} 
para um estado $|x_1 \ldots x_n\rangle$ da base computacional. Além disso, $H_C$ possui a forma genérica
\begin{equation}
H_C = \sum_{i,j=1}^n Q'_{ij}\,Z_i \otimes Z_j + \sum_{i=1}^n L'_i\,Z_i + K,
\label{eq:HamiltonianaQUBO}
\end{equation}
onde os novos coeficientes são obtidos dos coeficientes originais a partir da expansão de $\hat{x}_i$ em termos de $Z_i$. Em implementações práticas, o termo constante $K$ pode ser ignorado, uma vez que ele não altera o autoestado que minimiza o autovalor de $H_C$.

Como todos os termos em $H_C$ comutam entre si, a unitária associada $U_C$ pode ser expandida em um produto de unitárias associadas a cada termo, em qualquer ordem desejada. Os termos ``lineares'', envolvendo apenas operadores na base $z$ para um único qubit, são imediatamente identificados como rotações em torno do eixo $z$ da esfera de Bloch para esse qubit, que costumam ser facilmente implementáveis em plataformas de computação quântica remota já disponíveis atualmente \cite{Qiskit}. Já para os termos envolvendo operadores sobre dois qubits, nos valemos da identidade
\begin{equation}
\exp(-i k Z_i \otimes Z_j) = \text{CNOT}_{ij} \exp(-i k Z_j)\text{CNOT}_{ij},
\label{eq:ident_ZZ_CNOT}
\end{equation}
onde $\text{CNOT}_{ij}$ é a porta não-controlada (Eq.\ \eqref{eq:CNOT}) com controle no qubit $i$ e atuando sobre o qubit $j$. Essas portas também são implementáveis nessas plataformas de computação quântica, apesar de serem ainda bastante ruidosas e terem a conectividade entre os qubits ainda consideravelmente limitadas (isto é, nem todos os pares $i$ e $j$ são conectados diretamente, o que envolve a aplicação de mais portas CNOT para realizar uma conexão indireta). 

%{\color{red}[Acho que esse parágrafo pode ser removido, dado que agora tem a explicação mais detalhada abaixo.]} Sendo assim, para versões limitadas de um problema QUBO, o Hamiltoniano de custo pode ter sua evolução implementada na prática usando um modelo de circuito com portas quânticas. Uma ilustração do circuito correspondente pode ser vista na Fig. \ref{fig:QAOA_HBeHC} (b).

Na Fig. \ref{fig:QAOA_HBeHC}(a) mostramos as portas lógicas necessárias para construir a operação $U_B$. Nela aplicam-se rotações em todos os qubits no eixo $x$ sob um ângulo $2\beta_k$. Ou seja, ocorre uma rotação na esfera de Bloch de cada um dos qubits, essa rotação acontece como representado na Porta X da Fig. \ref{fig:EsferaBloch} mas o ângulo de rotação é de $2\beta_k$. O subíndice $k$ representa o contador de cada iteração.
Já a Fig. \ref{fig:QAOA_HBeHC}(b) mostra a implementação de $U_C$, aqui é usado o modelo estabelecido para a Hamiltoniana $H_C$ na Eq. \eqref{eq:HamiltonianaQUBO}. Para tal implementação, aplicam-se rotações no eixo $z$ sob ângulos de $2\gamma_{k}Q'_{1j}$ onde os coeficientes $Q'_{1j}$ são determinados individualmente em relação às operações entre os qubits $1$ e $j$. Antes e depois das rotações $R_z$, aplicam-se portas CNOT, como demonstrado na Eq. \ref{eq:ident_ZZ_CNOT}. Por fim aplicam-se  rotações, também no eixo $z$, com ângulos $2\gamma_{k}L'_{j}$. Observe que estas operações ocorrem entre pares de qubits sempre envolvendo o qubit $q_1$ (qubit controle) e os outros qubits $q_j$ (qubit alvo).

Note que os valores de $Q'_{ij}$ e $Q'_{ji}$ podem ser agregados em um só coeficiente, uma vez que $Z_i$ e $Z_j$ comutam entre si. Também é possível incorporar termos diagonais nos termos lineares, no problema clássico original, uma vez que variáveis binárias satisfazem $x_i^2 = x_i$. Usamos também a identidade \eqref{eq:ident_ZZ_CNOT} para implementar as unitárias associadas aos termos quadráticos.

A vantagem quântica do algoritmo QAOA foi investigada na Ref. \cite{farhi2019quantum}, onde foi mostrado  que a amostragem eficiente da distribuição de saída de QAOA, mesmo para o caso de menor profundidade de $p = 1$, implica no colapso da hierarquia polinomial, um forte indício de que não haja uma simulação clássica eficiente deste algoritmo.

De um modo geral, métodos variacionais, como o QAOA, fornecem limites superiores para a energia mais baixa de uma função custo com forma/simetria mapeada em um Hamiltoniano. Contudo, vale dizer, que a relativa aceleração do algoritmo quântico e a garantia de convergência são questões de pesquisa ainda em aberto. Ou seja, é possível que novos algoritmos clássicos mais eficientes sejam descobertos, bem como classe de problemas onde a convergência dos métodos quânticos é desafiada. Uma desvantagem conhecida dos métodos variacionais, famosa no domínio da química quântica, é a não extensividade da solução com o tamanho do sistema. Dito de outro modo, a energia calculada usando tais ferramentas pode não escalar com o tamanho do sistema  \cite{pople}. Porém, no domínio da otimização combinatória, este fato não é particularmente preocupante, uma vez que escalar o problema (acrescentar novos ativos, por exemplo), em geral, significa ter um novo sistema e todo cálculo precisa ser necessariamente refeito. Muito embora as limitações do método, há uma projeção de que uma demonstração de supremacia quântica  baseada no QAOA assim que um computador quântico de $420$ qubits (com correção de erros) esteja disponível \cite{dalzell}.

\subsection{Relação do QAOA com a Computação Adiabática}
\label{subsec:QAOA_adiabatico}

A conexão do QAOA com a computação adiabática se dá pela aplicação repetida dos operadores de custo e do misturador, que podem ser entendidas como uma discretização de uma evolução adiabática contínua.
A forma geral do problema adiabático é descrito na Eq. \eqref{eq:Hamiltonianogeral}, relembramos aqui a equação já com os Hamiltonianos inicial e final pertinentes ao problema do QAOA:
\begin{equation}
	H(t) = \left(1 - \frac{t}{T}\right)H_B + \left(\frac{t}{T}\right)H_C.
\label{eq:QAOAHamiltonianogeral}
\end{equation}
O Hamiltoniano do misturador $H_B$ será nosso Hamiltoniano inicial e o Hamiltoniano de custo $H_C$ será nosso Hamiltoniano final. Na formulação adiabática, a evolução é contínua e $t$ é o tempo transcorrido desde o começo da evolução, $T$ é o tempo final de evolução entre $H_B$ e $H_C$, isto é $t \in [0,T]$.

Notemos então que, de acordo com o teorema adiabático, para $T$ suficientemente grande (de modo que $H(t)$ varie lentamente), um autoestado de $H_B$ seria continuamente transformado no autoestado correspondente de $H_C$. Se o problema de otimização que queremos resolver com $H_C$ é um problema de minimização, começamos com o autoestado de menor autovalor de $H_B$. Analogamente, podemos começar com o autoestado de maior autovalor de $H_B$, como o estado inicial do QAOA, $\ket{\psi_{\textrm{ini}}}$, descrito na Eq. \eqref{eq:superposicao}, para resolvermos um problema de maximização. Podemos também começar com o mesmo estado e substituírmos $H_B$ por $-H_B$ no Hamiltoniano $H(t)$ para permanecermos com um problema de minimização.

Acima mostramos que o QAOA é um algoritmo projetado para ser implemetado em um computador quântico no modelo de circuito. No entanto, as portas quânticas são construídas realizando a evolução de um estado quântico a partir de um Hamiltoniano tal que $U(\alpha,H)=e^{-i\alpha H}$.
No caso da evolução de $H(t)$, dada $U(T,t_0)$ uma matriz que descreve a evolução de um estado quântico entre os tempos $t_0$ e $T$, podemos discretizar a evolução tal que
\begin{equation}
    U(T,t_0) = U(T,t_m)U(t_m,t_{m-1})...U(t_1,t_0),
\label{eq:U_discreto}
\end{equation}
onde $m$ é o número de divisões da evolução, ou seja, número de passos de tempo.  Conforme $m$ aumenta, a aproximação se torna mais precisa e cada passo da evolução passa a envolver um intervalo de tempo cada vez menor. Com um número suficientemente grande para $m$, cada termo da discretização pode ser bem aproximado por termos de primeira ordem no tempo, de modo que podemos reescrever a Eq. \eqref{eq:U_discreto} como uma função de diversas exponenciais:
\begin{equation}
    U(T,t_0)\approx e^{-\frac{i}{h} H_n(T-t_n)} e^{-\frac{i}{h} H_{n-1}(t_n-t_{n-1})} ... e^{-\frac{i}{h} H_{0}(t_1-t_{0})}.
\label{eq:U_exp_discreto}
\end{equation}

Podemos também separar a evolução de cada Hamiltoniano em evoluções de termos individuais. Como os Hamiltonianos $H_B$ e $H_C$ não comutam entre si, não podemos simplesmente separar a evolução deles em termos individuais, já que para operadores $A$ e $B$ que não comutam, vale que $e^{A+B}\neq e^Ae^B$.
Podemos, no entanto, decompor tal expressão usando a decomposição de Suzuki-Trotter \cite{suzuki1976generalized,suzuki1976relationship,suzuki1977convergence,suzuki1985decomposition}:
\begin{equation}
e^{A+B}= \lim_{p\to\infty}\left(e^{A/p}e^{B/p}\right)^p.
\label{eq:troterizacao}
\end{equation}
A evolução final aproximada pode ser escrita então como um produto alternado de exponenciais de $H_B$ e $H_C$, avaliadas em instantes diferentes. Isso dá o seguinte aspecto ao operador de evolução total
\begin{equation}
    U(t,t_0)\approx e^{-i \beta'_n H_B}e^{-i \gamma'_n H_C}\ldots e^{-i \beta'_1 H_B} e^{-i \gamma'_1 H_C},
\label{eq:U_exp_total_adiab}
\end{equation}
que possui exatamente a forma da expressão \eqref{eq:EstadoFinal_QAOA}, sendo que agora os parâmetros $\beta'$ e $\gamma'$ englobam a duração de cada evolução e o fator da decomposição de Suzuki-Trotter.

A importância da Eq. \eqref{eq:U_exp_total_adiab} é a de mostrar que, para um número de iterações do algoritmo QAOA $p \rightarrow \infty$, existem parâmetros $\beta'$ e $\gamma'$, pelo menos em princípio, que permitiriam o algoritmo simular com exatidão uma evolução adiabática. Com isso, e usando que a convergência ao valor ótimo do problema de otimização é garantido pelo teorema adiabático, obtemos uma garantia de que uma otimização adequada dos parâmetros $\beta$ e $\gamma$ do QAOA deve também permitir convergência ao valor ótimo do problema.
Em outras palavras, podemos sempre encontrar um $p$ e um conjunto de ângulos $\gamma_k$ e $\beta_k$ que tornam $F_p$ (Eq. \eqref{eq:OtimoQAOA}) o mais próximo do menor nível de energia do espectro em questão \cite{qaoa}.

Apesar dessa condição ser suficiente para garantir a convergência, nada implica que ela seja necessária. Consequentemente, o QAOA pode ser mais eficiente que a computação adiabática, uma vez que, através da otimização dos parâmetros, possivelmente fugindo da prescrição dada pela evolução adiabática, podemos encontrar boas aproximações para o problema de otimização com poucas iterações do QAOA.

\subsection{Aplicação à Otimização de Portfólio}
\label{subsec:QAOA_aplicacao}

Para a abordagem quântica do problema de otimização de portfólio, ao invés de pesar as diferentes ações, determinaremos quais ações incluir e quais excluir de uma carteira, usando método do QAOA para a tarefa de otimização de portfólio \cite{hodson2019portfolio}. É, assim, um problema relacionado ao problema clássico, porém distinto.

Conforme a Eq. \eqref{eq:OtimizacaoRestrita}, além dos ingredientes usados na abordagem clássica (os retornos e a matriz de covariância), precisamos informar alguns parâmetros. Um deles é o apetite ao risco $b$, que fixamos em $b = 0.5$, correspondente ao maior risco observado na abordagem clássica. O outro é o parâmetro $B$, que representa o orçamento, neste caso quantas ações colocar na carteria. Como utilizamos as mesmas ações anteriores e sabendo que o portfólio com máximo Sharpe Ratio na abordagem clássica se concentra majoritariamente em duas ações, fixamos $B=2$ com um intuito de contrastar com os resultados anteriores. Observe que o termo de restrição provoca uma penalização ao construir um portfólio com um número de ações diferente de $B$, ver Ref. \cite{noteRBEF}. 

Dito de outro modo, ao fixar $B=2$ pretendemos responder a seguinte pergunta: será que o QAOA retornará como portfólio ideal uma carteira semelhante à carteria clássica? Ou seja, uma carteira com Braskem e Vale, tal qual obtido anteriormente? Observe que as carteiras com melhores Sharpe Ratio na Tabela \ref{tab:port} são concentradas praticamente em proporções iguais em BRKM5 e VALE3. Veja também a Ref. \cite{noteRBEF} para mais detalhes.

Assim, usando a Tabela \ref{tab:matriz_cov} podemos escrever a função objetiva (relacionada ao Hamiltoniano) a ser minimizada usando o QAOA como
\begin{eqnarray}
    C(x) = -x_{1}-x_{2}-x_{3}-x_{4}+(0.37x_{1}x_{2})+(0.28x_{1}x_{3}) \nonumber \\
    (0.33x_{1}x_{4})+(0.18x_{2}x_{3})+(0.36x_{2}x_{4})+(0.24x_{3}x_{4}) \nonumber \\ - (2 -x_1-x_2-x_3-x_4)^2, \nonumber \\
\label{eq:exemplo_QUBO}
\end{eqnarray}
onde $x_1$, $x_2$, $x_3$ e $x_4$ equivalem, respectivamente, a BRKM5, ITUB4, KLBN e VALE3, sendo variáveis binárias, ou seja, podem assumir valores de $0$ ou $1$. Observe como o termo de de restrição penaliza carteiras com número de ações diferentes do orçamento ($B=2$). Por exemplo, verifique o valor da função objetiva para a escolha das $4$ ações ou nenhuma delas em Ref. \cite{noteRBEF}, visto que escolher exatamente duas ações anula a penalidade.

\begin{figure}[t!]
\centering
\includegraphics[width=0.475\textwidth]{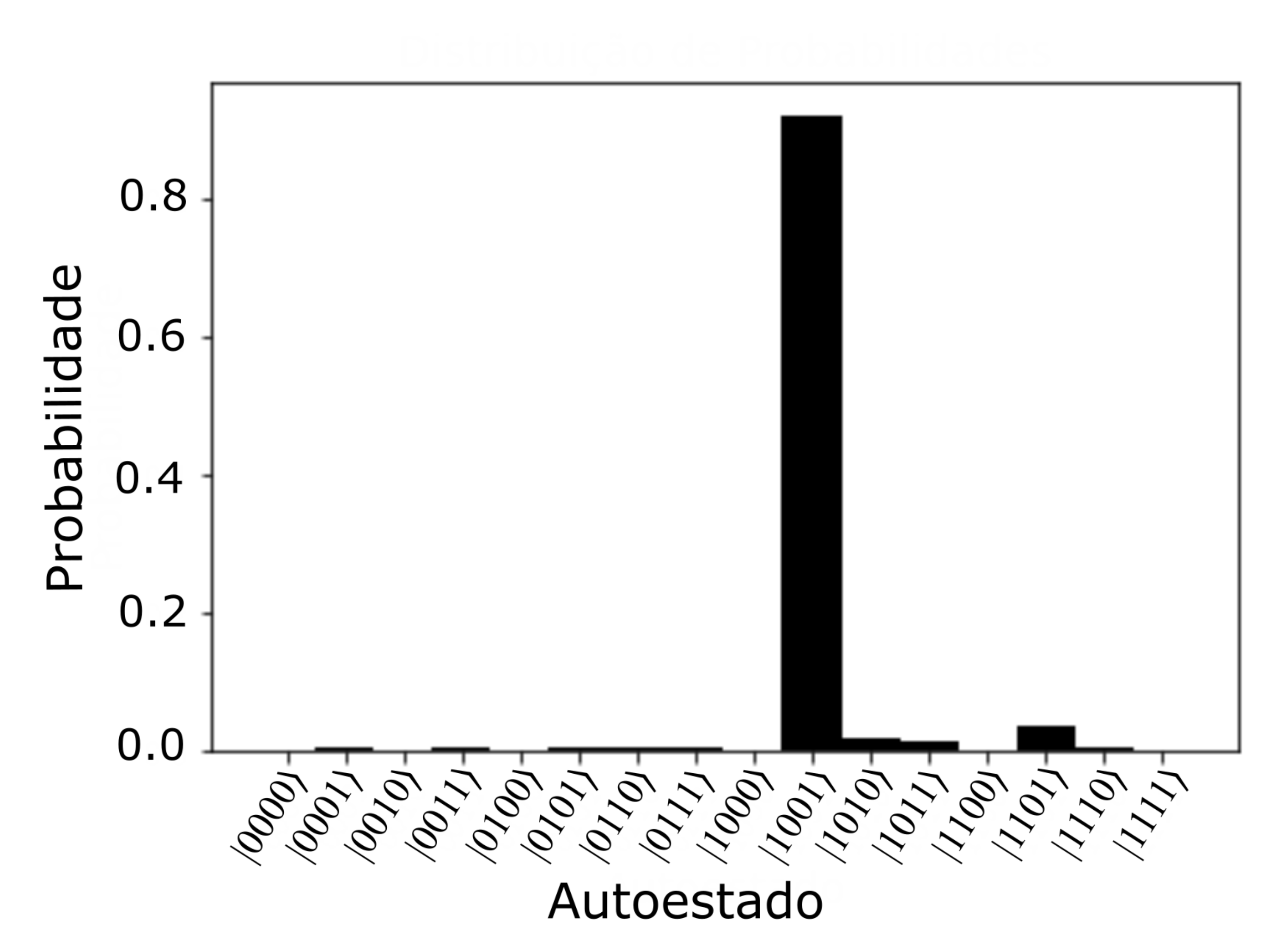}
\caption{Histograma da distribuição de probabilidades de medição de cada estado da base computacional, considerando-se a solução ótima obtida pelo QAOA com $20$ camadas.}
\label{fig:hist}
\end{figure}

Por meio do QAOA, encontramos, para cada uma das $2^4 = 16$ combinações possíveis das variáveis binárias, ou seja, para cada um dos vetores de estado ($\ket{0000}, \ket{0001}, ..., \ket{1110}, \ket{1111}$), um valor para a energia associada e a probabilidade de o respectivo vetor ser a solução ideal. Como vimos na Seção \ref{sec:QAOA}, o QAOA permite transformar o estado de superposição uniforme no autoestado  correspondente à menor energia por meio da otimização dos parâmetros {$\beta$, $\gamma$}. 

Na Fig. \ref{fig:hist} mostramos que, ao utilizar o QAOA com $20$ camadas a solução ótima encontrada, sendo aquela com menor energia associada $H_{\text{min}}$,  representa o vetor $\ket{1001}$, que corresponde a uma carteira com incluindo Braskem ($x_1 =1$) e Vale ($x_4 = 1$) e excluindo Itaú ($x_2 = 0$) e Klabin ($x_2 = 0$). Note que a probabilidade máxima, de quase $90\%$, é justamente aquela associada à solução do problema, o autoestado $\ket{1001}$, aquele com a menor energia para o Hamiltoniano considerado.

\begin{figure}[t!]
\centering
\includegraphics[width=0.475\textwidth]{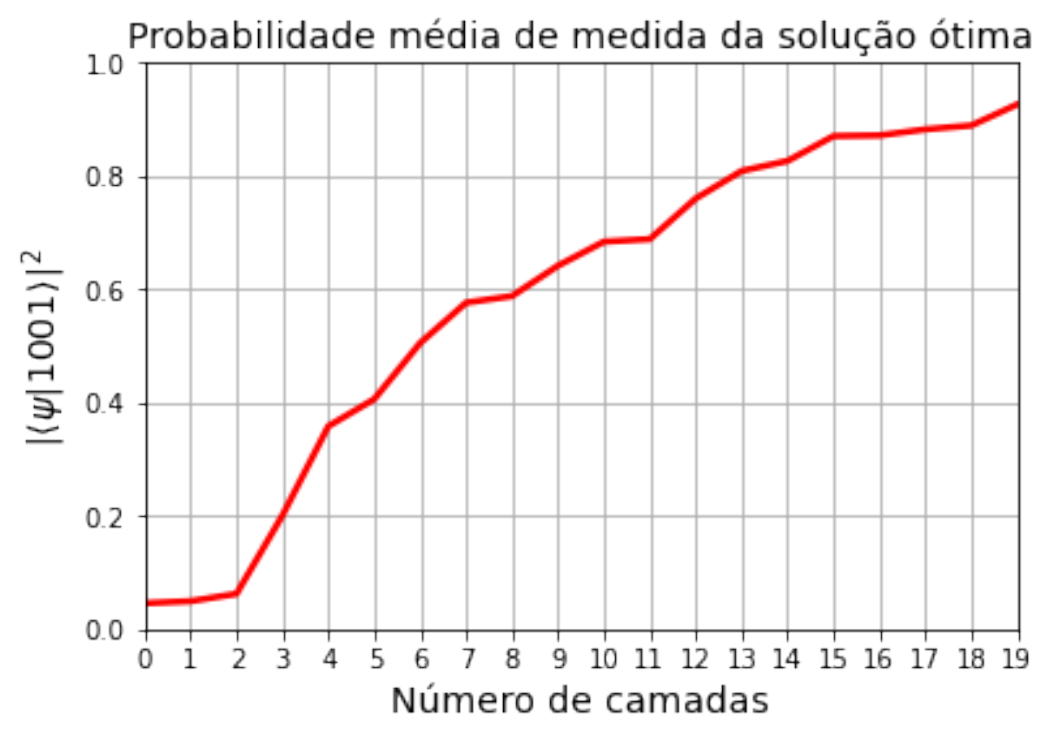}
\caption{Probabilidade $\vert\langle \psi(\beta,\gamma)  \vert 1001 \rangle\vert ^2$ de que o estado encontrado pelo QAOA seja encontrado na solução ótima $\ket{1001}$, plotada em função do número de camadas no QAOA. Para cada número de camadas, consideramos uma média sobre as duas melhores dentre $10$ otimizações.}
\label{fig:prob}
\end{figure}

Assim, o resultado da otimização combinatória de portfólio encontrado usando o QAOA é condizente com a carteira ótima encontrada com o método clássico, embora seja válido mencionar novamente que são problemas semelhantes, porém distintos. Ocorre, ainda, que o QAOA é um método heurístico, de modo que não sabemos, de antemão, o melhor valor para o número de camadas. Além disto, a convergência pode variar para execuções distintas do algoritmo. Deste modo, investigamos, na Fig. \ref{fig:prob}, como a probabilidade de que o estado $\ket{\psi(\beta,\gamma)}$ encontrado pelo QAOA seja medido no autoestado fundamental $\ket{1001}$ varia com número de camadas após tomar a média das duas melhores dentre $10$ execuções do código, ver Ref. \cite{noteRBEF}. Note que o comportamento não é monotônico, ou seja, aumentar o número de camadas nem sempre melhora a probabilidade associada ao vetor $\ket{1001}$ ser a solução ideal. 

\begin{figure}[t!]
\centering
\includegraphics[width=0.475\textwidth]{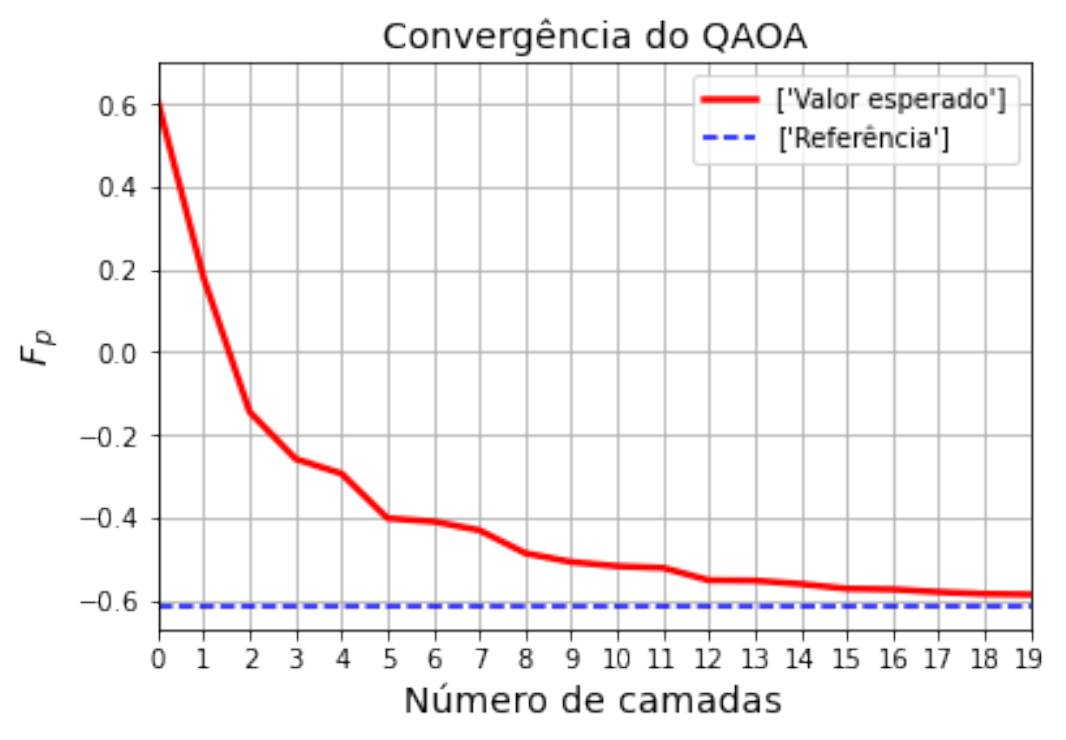}
\caption{A energia da solução $F_p$ (Eq. \eqref{eq:OtimoQAOA}) encontrada pelo QAOA como função do número de camadas. Como referência plotamos em azul a energia fundamental, associada ao autoestado $\ket{1001}$ do Hamiltoniano dado por $\langle H\rangle_{min}=-0,6165$. Para cada número de camadas, consideramos uma média sobre as duas melhores dentre $10$ otimizações.}
\label{fig:conv}
\end{figure}

A mesma investigação pode ser feita olhando-se para a energia da solução encontrada pelo QAOA e compará-la com o valor mímino associado ao vetor $\ket{1001}$, o que é apresentado na Fig. \ref{fig:conv}. Note que a energia tem um comportamento mais suave, porém também oscila mesmo tomando-se a média das duas melhores dentre $10$ diferentes otimizações para um dado número de camadas. Notamos um comportamento assintótico em torno de $16$ a $20$ camadas.

Cabe mencionar que a qualidade da solução obtida por meio do QAOA para um determinado problema depende da performance do otimizador clássico usado para otimizar os parâmetros variacionais. Deste modo, nossa análise com relação ao número de camadas não é a mais ampla possível. Entretanto, foge ao escopo desse artigo comparar as soluções obtidas com distintos otimizadores como, por exemplo, COBYLA, BOBYQA, Nelder-Mead, SLSQP, SPSA, entre outros \cite{optimizersQVA}. Fazendo uso do Jupyter Notebook disponível na Ref. \cite{noteRBEF}, pode-se realizar extensivas simulações e, até mesmo, adaptar livremente o código para problemas de seu interesse. 

Para este artigo, o QAOA foi implementado através do ATOS Quantum Learning Machine (QLM) \cite{myqlm} que é um ambiente de teste para a computação quântica desenvolvida pela empresa francesa ATOS e adquirido pelo SENAI CIMATEC para o desenvolvimento de algoritmos quânticos no Centro de Computação Quântica da América Latina. O dispositivo é capaz de simular até 35 qubits e inclui a simulação de ruídos, emulando a arquitetura de computadores quânticos reais. A escolha por trazer uma solução de certo modo completamente nacional visa fortalecer as colaborações e pesquisas científicas em tecnologias quânticas.

O QLM também possui interoperabilidade com outras frameworks de computação quântica, o que possibilita o uso de computadores quânticos através da plataforma. Além disso, o myQLM, framework disponibilizado gratuitamente, possibilita também o desenvolvimento e simulação de algoritmos quânticos por usuários domésticos, podendo ser instalado localmente. Por sua completeza, a utilização do QLM para o desenvolvimento de algoritmos de computação quântica se torna interessante, podendo ser utilizado em diferentes etapas do desenvolvimento de algoritmos quânticos. Algumas outras vantagens são: a existência de plug-ins especiais para realizar métodos variacionais orientados a era NISQ (como VQE, QAOA), bem como uma interface para conectar-se com os processadores quânticos disponíveis e as principais estruturas de programação quântica, tais como Qiskit, ProjectQ, Cirq etc. \footnote{Para instalação do myQLM recomendamos seguir a própria documentação do software, disponível em: \href{https://myqlm.github.io/myqlm_specific/install.html}{https://myqlm.github.io/myqlm\_specific/install.html}. Na nossa documentação, também apresentamos um roteiro de como fazer a instalação: \href{https://github.com/askery/computacao-quantica-aplicada-ao-mercado-financeiro}{https://github.com/askery/computacao-quantica-aplicada-ao-mercado-financeiro}}.

\section{Conclusão}
\label{sec:conclusao}

Após um longo período sendo apenas uma área acadêmica, mais notadamente nas fronteira da Física, a computação quântica tem atraído crescente atenção de governos, empresas e indústrias por suas promessas tecnológicas. Mesmo que um computador quântico com correção de erros ainda não seja uma realidade, o panorama traçado por empresas do setor mostram que tal feito pode ser alcançado. Em seu \emph{roadmap} \cite{ibmroadmap}, a IBM aposta em chips quânticos com mais de 1000 qubits até 2023 e uma plataforma escalável atingindo 1 milhão de qubits  até o final desta década. Em contrapartida, a IonQ planeja um computador quânticos com mais de 1000 qubits \cite{ionq}, já com correção de erros, até 2028. Razão mais que suficiente para que começemos a nos preparar e entendamos como estas máquinas quânticas de um futuro próximo poderão ser operadas.

Dentro deste contexto, o objetivo deste artigo foi o de apresentar um algoritmo de destaque na literatura, o QAOA, \textit{Quantum Approximation Optimzation Algorithm}, e uma de suas aplicações em um importante problema financeiro, a otimização de portfólios. Não só expusemos de forma concisa os principais conceitos, mas também fornecemos provas de princípio desta aplicação, considerando um problema simples com $4$ ativos disponíveis na bolsa brasileira. Como forma de motivação, disponibilizamos um Jupyter Notebook com explicações detalhadas do mercado financeiro e dos algoritmos tanto clássico quanto quânticos que foram utilizados.

A segunda revolução quântica, com dispositivos tecnológicos que exploram toda a capacidade informacional da mecânica quântica, já é uma realidade. Esperamos que este artigo possa impulsionar ainda mais esta área em nosso país, que apesar de grande qualidade acadêmica no tema, ainda carece de uma maior integração com setores privados da sociedade. A ciência quântica promete moldar o desenvolvimento de tecnologias do século 21. Ainda é tempo para que o Brasil reassuma sua posição de destaque neste tema.

\section*{Agradecimentos}
Agradecemos ao Instituto Serrapilheira (Grant n.
Serra-1708-15763), a John Templeton Foundation via o projeto Q-CAUSAL No 61084, a Simons Foundations (Grant Number 884966, AF), ao Conselho Nacional de Desenvolvimento Científico e Tecnológico (CNPq) por meio do Instituto Nacional de Ciência e Tecnologia de Informação Quântica (INCT-IQ) e das Bolsas: (N. 423713 / 2016-7, N. 307172 / 2017-1, N. 406574 / 2018-9 e N. 307295/2020-6), ao Ministério da Ciência, Tecnologia e Inovação do Brasil,
e Comunicações (MCTIC) e ao Ministério da Educação (MEC). A.C. agradece à Universidade Federal de Alagoas (UFAL) por um licença de cooperação científica na Universidade Federal do Rio Grande do Norte (UFRN).

\bibliography{refs.bib}
\bibliographystyle{ieeetr}

\end{document}